\documentclass{LMCS}

\usepackage{enumerate}
\usepackage{hyperref}
\usepackage{bussproofs}
\usepackage{stmaryrd}
\usepackage{pstricks,pst-tree,pst-node}
\usepackage{cmll}
\usepackage{amssymb}
\usepackage{amsmath}
\usepackage{latexsym}
\usepackage{graphicx}
\usepackage{proof}

\newcommand{\buparrowd}{\pmb{\mathord{\uparrow}}}
\newcommand{\bdownarrowd}{\pmb{\mathord{\downarrow}}}
\newcommand{\uparrowd}{\mathord{\uparrow}}
\newcommand{\downarrowd}{\mathord{\downarrow}}
\newcommand{\Sum}{\textstyle\sum}
\newcommand{\Bigwedge}{\textstyle\bigwedge}
\newcommand{\nondet}{-\!\!\!\!\!\rightharpoonup}

\def\llone{\mbox{\bf 1}}
\def\llwith{\mathbin{\rm\&}}
\def\lltensor{\otimes}

\def\llplus{\oplus}
\newcommand{\llpar}{\,\rotatebox[origin=c]{180}{\&}\,}
\newcommand{\llzero}{\mbox{\bf 0}}

\newcommand{\bA }{{\mathbf{A}}}

\newcommand{\bD }{{\mathbf{D}}}

\newcommand{\bG }{{\mathbf{G}}}

\newcommand{\bK }{{\mathbf{K}}}
\newcommand{\bL }{{\mathbf{L}}}

\newcommand{\bX }{{\mathbf{X}}}
\newcommand{\bY }{{\mathbf{Y}}}

\newcommand{\bN }{{\mathbf{N}}}
\newcommand{\bM }{{\mathbf{M}}}
\newcommand{\bP }{{\mathbf{P}}}
\newcommand{\bQ }{{\mathbf{Q}}}
\newcommand{\bR }{{\mathbf{R}}}

\newcommand{\bGamma }{{\mathbf{\Gamma}}}
\newcommand{\bDelta }{{\mathbf{\Delta}}}
\newcommand{\bLambda }{{\mathbf{\Lambda}}}

\newcommand{\bPsi }{{\mathbf{\Psi}}}
\newcommand{\bTheta }{{\mathbf{\Theta}}}

\newcommand{\cA }{{\mathcal{A}}}

\newcommand{\cD }{{\mathcal{D}}}

\newcommand{\cG }{{\mathcal{G}}}

\newcommand{\cN }{{\mathcal{N}}}
\newcommand{\cM }{{\mathcal{M}}}
\newcommand{\cP }{{\mathcal{P}}}
\newcommand{\cV }{{\mathcal{V}}}

\newcommand{\cR }{{\mathcal{R}}}

\newcommand{\sD }{{\mathsf{D}}}

\newcommand{\sN }{{\mathsf{N}}}
\newcommand{\sM }{{\mathsf{M}}}
\newcommand{\sL }{{\mathsf{L}}}
\newcommand{\sP }{{\mathsf{P}}}
\newcommand{\sQ }{{\mathsf{Q}}}
\newcommand{\sR }{{\mathsf{R}}}

\newcommand{\sGamma}{{\mathsf{\Gamma}}}

\newcommand{\sPl }{{\mathsf{p}}}
\newcommand{\sNl }{{\mathsf{n}}}
\newcommand{\sMl }{{\mathsf{m}}}

\newcommand{\vx }{\vec{x}}
\newcommand{\vy }{\vec{y}}

\newcommand{\vz }{\vec{z}}

\renewcommand{\b}{{}^{\bot}}
\newcommand{\coa }{{\overline{a}}}
\newcommand{\cob }{{\overline{b}}}
\newcommand{\coc }{{\overline{c}}}
\newcommand{\coal }{{\overline{\alpha}}}
\newcommand{\cobe }{{\overline{\beta}}}

\newcommand{\one}{\mathbf{1}}
\newcommand{\zero}{\mathbf{0}}
\newcommand{\toppo}{\pmb{\top}}
\newcommand{\botto}{\pmb{\bot}}

\newcommand{\pl}{ \llbracket}
\newcommand{\pr}{ \rrbracket}

\newcommand{\nv}{\not\!\!\!\bot\nobreak}

\theoremstyle{plain}
\def\eg{{\it e.g.}, }
\def\cf{{\it cf.} }
\def\ie{{\it i.e.}, }

\newcommand{\Bra}[1]{\langle #1 \rangle}
\newcommand{\Norm}[1]{[\![ #1 ]\!]}
\newcommand{\Ov}[1]{\overline{#1}}
\newcommand{\Rel}{\; \cR\; }

\def\doi{6 (4:11) 2010}
\lmcsheading%
{\doi}
{1--35}
{}
{}
{Nov.~22, 2009}
{Dec.~22, 2010}
{}

\begin{document}

\title[On the meaning of logical completeness]{On the meaning of logical completeness}

\author[M.~Basaldella]{Michele Basaldella\rsuper a}	
\address{{\lsuper{a,b}}Research Institute for Mathematical Sciences, Kyoto University, Kitashirakawa Oiwakecho, Sakyo-ku, Kyoto 606-8502, Japan.}	
\email{\{mbasalde,terui\}@kurims.kyoto-u.ac.jp}  
\thanks{{\lsuper a}Supported by JSPS Postdoctoral Fellowship Program for Foreign
Researchers grant 2008803.}

\author[K.~Terui]{Kazushige Terui\rsuper b}	
\address{\vskip-6 pt}	
\thanks{{\lsuper{a,b}}This work was supported by JSPS KAKENHI 21700041.}	






\begin{abstract}
  \noindent G\"odel's completeness theorem is concerned with provability,
while Girard's theorem in ludics (as well as
full completeness theorems in game semantics)
are concerned with proofs.
Our purpose is to look for a connection between these two
disciplines. Following a previous work \cite{Basaldella09},
we consider an extension of the original ludics with contraction
and universal nondeterminism, which play dual roles,
in order to capture a polarized
fragment of linear logic and thus a constructive variant of
classical propositional logic.

We then prove a completeness theorem for proofs in this extended setting:
for any behaviour (formula) $\bA$ and any design (proof attempt) $P$,
either $P$ is a proof of $\bA$ or there is a model $M$ of $\bA^{\bot}$
which defeats $P$.
Compared with proofs of full completeness in game semantics, ours
exhibits a striking similarity with proofs of G\"odel's completeness,
in that it explicitly constructs a countermodel essentially using
K\"onig's lemma, proceeds by induction on formulas, and implies an
analogue of L\"owenheim-Skolem theorem.
\end{abstract}

\maketitle

\section*{Introduction}\label{S:one}
\noindent G\"odel's completeness theorem (for first-order classical
logic) is one of the most important theorems in logic. It is concerned
with a duality (in a naive sense) between proofs and models:
For every proposition $\bA$,
\begin{center}
{\em either\quad $\exists P (P\vdash \bA)$\quad or\quad $\exists M (M \models \neg \bA)$.}
\end{center}
Here $P$ ranges over the set of proofs, $M$ over the class of models,
and $P\vdash \bA$ reads ``$P$ is a
proof of $\bA$.'' One can imagine a debate on a general proposition
$\bA$, where Player tries to justify $\bA$ by giving a proof
and Opponent tries to refute it by giving a countermodel.
The completeness theorem states that exactly one of them wins.
Actually, the theorem gives us far more insights than stated.
\begin{enumerate}[\hbox to8 pt{\hfill}]
\item\noindent{\hskip-12 pt\bf Finite proofs vs infinite models:}\
A very crucial point is that proofs are always finite,
while models can be of arbitrary cardinality. 
Completeness thus implies compactness and
L\"owenheim-Skolem  theorems, leading to
constructions of various nonstandard models.

\item\noindent{\hskip-12 pt\bf Nondeterministic principles:}\
{\em Any} proof of G\"odel's completeness theorem
relies on a strong nondeterministic principle such as K\"onig's
or Zorn's lemma, in contrast to the trivial completeness theorem
with respect to the class of boolean algebras.

\item\noindent{\hskip-12 pt\bf Matching of two inductions:}\
\emph{Provability} is defined by induction on proofs, while
\emph{truth}  by induction on formulas.
The two inductions
are somehow ascribed to the essence of syntax and semantics, respectively,
and the completeness theorem states that they do match.
\end{enumerate}

\medskip\noindent Unlike the real debate, however, there is no interaction between proofs and
models in G\"odel's theorem.
A more interactive account of completeness is given
by Girard's {\em ludics} (\cite{Girard99b,DBLP:journals/mscs/Girard01}; see
\cite{DBLP:journals/tcs/Faggian06,DBLP:journals/corr/abs-cs-0501039}
for good expositions). Ludics is a variant of game semantics,
which has the following prominent
features.
\begin{enumerate}[\hbox to8 pt{\hfill}]
\item\noindent{\hskip-12 pt\bf Monism:}\ Proofs and models are
not distinguished by
their ontological status, but by their structural properties.
The common objects are
called {\em designs}.
\item\noindent{\hskip-12 pt\bf Existentialism:}\ {\em Behaviours} (semantic types)
are built from designs, in contrast to the ordinary game semantics
(\eg Hyland-Ong \cite{DBLP:journals/iandc/HylandO00}) where
one begins with the definition of arenas (types)
and then proceeds to strategies (proofs).
\item\noindent{\hskip-12 pt\bf Normalization as interaction:}\ Designs (hence proofs and models) interact
together via normalization.
It induces
an {\em orthogonality relation} between designs
in such a way that $P \bot M$
 holds if the normalization of $P$ applied to $M$ converges.
A behaviour $\bA$ is defined to be a set of designs which is equivalent to
its biorthogonal ($\bA= \bA^{\bot\bot}$).
\end{enumerate}

\medskip\noindent In this setting,
Girard shows a completeness theorem for proofs
 \cite{DBLP:journals/mscs/Girard01}, which roughly
claims that any ``winning''
design in a behaviour is a proof of it.
In view of the interactive definition
of behaviour, it can be rephrased as follows:
For every (logical) behaviour $\bA$ and every (proof-like) design $P$,
\begin{center}
{\em either\quad $P\vdash \bA$\quad  or\quad $\exists M(M \models \bA^\bot$ and $M$
defeats $P$).}
\end{center}
Here, ``$M\models \bA^\bot$'' means $M \in \bA^\bot$, and
``$M$ defeats $P$'' means $P \nv M$.
Hence 
the right disjunct is equivalent to
$P\not\in
\bA^{\bot\bot} = \bA$.
Namely, $P \in \bA$ if and only if $P\vdash \bA$,
that is a typical full completeness statement.
Notice that $M\models \bA^\bot$
no more entails absolute unprovability
of $\bA$ (it is rather relativized to each $P$), and there is
a real interaction between proofs and models.

Actually, Girard's original ludics is so limited that it corresponds to
a polarized fragment of multiplicative additive linear logic, which is
too weak to be a stand-alone logical system. As a consequence,
one does not really observe
an opposition between
finite proofs and infinite models, since one can always assume that the countermodel
$M$ is finite (related to the finite model property for $\mathbf{MALL}$
\cite{DBLP:journals/jsyml/Lafont97}).
Indeed, proving the above completeness
is easy once \emph{internal completeness} (a form
of completeness which does not refer to any proof system \cite{DBLP:journals/mscs/Girard01})
 for each logical connective
has been established.

In this paper, we employ a term syntax for designs introduced in \cite{Terui08}, and
extend Girard's  ludics with duplication (contraction)
and its dual: universal nondeterminism (see \cite{Basaldella09}
and references therein).
Although our term approach disregards some interesting
 locativity-related phenomena
(\eg normalization as merging of orders and
different sorts of tensors
 \cite{DBLP:journals/mscs/Girard01}),
our calculus is easier to manipulate and closer to the  tradition
of $\lambda$, $\lambda\mu$,  $\overline{\lambda} \mu\tilde{\mu}$,
 $\pi$-calculi and other more recent syntaxes for focalized classical logic (\eg \cite{curmac}).
Our resulting framework is as strong as
a polarized fragment of linear logic with exponentials
(\cite{DBLP:journals/corr/abs-cs-0501039}; see also
\cite{DBLP:journals/apal/Laurent04}), which is in turn as strong as
a constructive version of classical propositional logic.

We then prove
the completeness theorem above in this extended setting.
Here, universal nondeterminism is needed on
the model side to well interact with duplicative designs on the proof side.
This is comparable to the need of ``noninnocent" (and sometimes even nondeterministic)
Opponents to have full completeness with respect to deterministic,
but nonlinear Player's strategies.
Unlike before, we cannot anymore assume the finiteness of
models, since they are not sufficient to refute infinite proof
attempts. As a result, our proof is nontrivial, even after
the internal completeness theorem has been proved.
Indeed,
our proof
exhibits a striking similarity with
Sch\"utte's proof of G\"odel's completeness theorem \cite{Schutte56}.
Given a (proof-like) design $P$ which is not a proof of $\bA$,
we explicitly construct a countermodel $M$ in $\bA^\bot$ which defeats $P$,
essentially
using K\"onig's lemma. Soundness is proved by induction on proofs, while
completeness is by induction on types.
Thus our theorem gives matching of
two inductions.
Finally, it implies an
analogue of L\"owenheim-Skolem theorem,
 (and also the finite model property for the linear fragment),
which well illustrates the opposition between finite proofs and
infinite models with arbitrary cardinality.

In game semantics, one finds a number of similar full completeness
results. However, the connection with G\"odel's completeness seems
less conspicuous than ours. Typically, innocent
strategies in Hyland-Ong games most naturally correspond to
{\em B\"ohm trees}, which can be {\em infinite}
(\cf \cite{Curiennote}). Thus, in contrast to our result,
one has to impose
finiteness/compactness on strategies
in an external way, in order
to have a correspondence with {\em finite} $\lambda$-terms.
Although this is also the case in \cite{Basaldella09}, we show that
such a finiteness assumption is not needed in ludics: infinitary
proof attempts are always defeated by infinitary models.

The paper is organized as follows. In Section \ref{s-prelim}
we describe  the syntax of (untyped) designs;  in Section \ref{s-behaviour}
we move to a typed setting and introduce
behaviours (semantic types). In Section \ref{s-full}
we introduce our proof system and prove completeness for proofs.
Finally,  Section \ref{s-conclusion} concludes the paper.

\section{Designs}\label{s-prelim}

\subsection{Syntax} \label{syntax}

In this paper, we employ
a process calculus notation for designs,
inspired by the close relationship
between ludics and linear $\pi$-calculus \cite{DBLP:conf/tlca/FaggianP07}.
Precisely, we extend the syntax
introduced by the second author  \cite{Terui08}  adding a (universal) nondeterministic choice operator $\Bigwedge$.

Although \cite{Terui08} mainly deals with linear designs,
its syntax is designed to
deal
 with nonlinear ones  without any difficulty.
However, in order to obtain \emph{completeness},
we also need to  incorporate
the dual of nonlinearity, that is {\em universal nondeterminism}
\cite{Basaldella09}.
It is reminiscent of differential linear logic
\cite{DBLP:journals/tcs/EhrhardR06}, which has
nondeterministic sum as the dual of contraction; the duality
is essential for the separation property
\cite{DBLP:conf/lpar/MazzaP07} (see also
\cite{Dezani-Intrigila-Venturini:ICTCS-98} for separation of B\"ohm trees).
A similar situation also arises
in Hyland-Ong game semantics
\cite{DBLP:journals/iandc/HylandO00},
where
{\em nonlinear} strategies for Player may contain a play
in which Opponent behaves {\em noninnocently}; Opponent's noninnocence is again
essential for full completeness.

Designs are built over a given
\emph{signature} $\cA=(A,\mathsf{ar})$, where
$A$ is a set of \emph{names} $a,b,c,\ldots$ and
$\mathsf{ar} : A \longrightarrow\mathbb{N}$ is
a function
which assigns to each name $a$ its \emph{arity} $\mathsf{ar}(a)$.
Let $\cV$ be a countable set of variables $\cV=\{x,y,z,\ldots\}$.

Over a fixed signature $\cA$,
a
{\em positive action} is $\coa$ with $a\in A$, and
a
{\em negative action} is $a(x_1, \dots, x_n)$
where variables $x_1, \dots, x_n$ are distinct and $\mathsf{ar}(a)= n$.
We often abbreviate a sequence of variables
$x_1,\ldots,x_n$ by $\vx$. In the sequel, we always
assume that an expression of the form
$a(\vec{x})$  stands for
a negative action, \ie $\mathsf{ar}(a) =n$ and $\vx$
is a sequence consisting of $n$ distinct variables.
If $a$ is a nullary name  we simply write
$a$ for the negative action on name $a$.


\begin{defi}[Designs] \label{designs}
For a fixed signature $\cA$,
the class of \textbf{positive designs} $P, Q, \dots$,
that of \textbf{predesigns} $S, T, \dots$, and
that of \textbf{negative designs} $N, M, \dots$
are \emph{coinductively} defined as follows:
$$
\begin{array}{rclc}
P  & ::=  &  \Omega & \mbox{ (\emph{divergence}), } \\
 &  \big| & \Bigwedge \{S_i : i\in I\} & \mbox{ (\emph{conjunction}), } \\
S & ::= &
   N_0 |\coa \langle N_1,\ldots,N_n \rangle & \mbox{ (\emph{predesign}), } \\
 N & ::= &   x  & \mbox{ (\emph{variable}), } \\
 & \big| &  \Sum a(\vx).P_a & \mbox{ (\emph{abstraction}), } \\
 \end{array}
$$
where:
\begin{enumerate}[$\bullet$]
\item $\mathsf{ar}(a) = n$;
\item $\vx = x_1, \ldots, x_n$ and
 the formal sum $\Sum a(\vx).P_a$ has
 $|A|$-many components $\{a(\vx).P_a\}_{a \in A}$;
\item
$\Bigwedge \{S_i: i\in I\}$ is built from a set
 $\{S_i: i\in I\}$ of predesigns with $I$ an arbitrary index set.
\end{enumerate}
We denote arbitrary designs by
$D,E,\ldots$.
The set of  \textbf{designs}, consisting of all positive, negative and predesigns, is denoted by
$\mathcal{D}$.
Any subterm $E$ of $D$ is equivalently called a \emph{subdesign} of $D$.
\end{defi}

\medskip\noindent Notice that designs are \emph{coinductively} defined objects.
In particular, infinitary designs are included in our syntax,
just as in the original ludics \cite{DBLP:journals/mscs/Girard01}.
It is strictly necessary, since
we want to express both  proof attempts and countermodels as designs,
both of which tend to be infinite.


Informally, designs may be regarded as infinitary
$\lambda$-terms with {\em named} applications, {\em named} and {\em superimposed} abstractions and a \emph{universal nondeterministic}
choice operator $\Bigwedge$.

More specifically, a predesign
$  N_0 |\coa \langle N_1,\ldots,N_n \rangle$  can be thought of as
iterated application $N_0 N_1 \cdots N_n =
(\cdots((N_0 N_1)N_2) \cdots N_n) $ of  an $n$-ary name $a\in A$.
In the sequel, we may abbreviate
$ N_0 |\coa \langle N_1,\ldots,N_n \rangle$
by   $  N_0 |\coa \langle \vec{N}\rangle$. If $a$ is a nullary
name we simply write $  N_0 |\coa$.

On the other hand, a negative design of the form
$a(\vx).P_a$ can be thought of as iterated abstraction
$\lambda\vx. P_a = \lambda x_1.(\lambda x_2 .(\cdots (\lambda x_n.P_a) \cdots))  $ of an $n$-ary name $a\in A$. A family
$\{a(\vx).P_a\}_{a\in A}$ of abstractions indexed by $A$
is then superimposed to form a
negative design $\Sum a(\vx).P_a$.
Since
$\Sum a(\vx).P_a$ is built from  a family
 indexed by  $A$, there cannot be
any overlapping of name in the sum.
Each $a(\vx).P_a$ is called an
(additive) {\bf component}.

A predesign is called a \textbf{cut} if it is
of the form
$(\Sum a(\vx).P_a)| \cob \langle N_1,\ldots, N_n \rangle$.
Otherwise, it is of the form
$x | \coa \langle N_1,\ldots, N_n \rangle$
and called a \textbf{head normal form}.

As we shall see  in  detail in Subsection  \ref{normalization subsection},
cuts have substantial  computational significance in our setting: in fact a cut $(\Sum a(\vx).P_a)| \cob \langle \vec{N} \rangle$ can be \emph{reduced} to another design
$
P_b [\vec{N}/\vy]$.
Namely, when the application is of name $b$, one picks up the component
$b(\vy).P_b$ from the family $\{a(\vx).P_a\}_{a\in A}$.
Notice  that the arities of $\vy$ and $\vec{N}$ always agree.
Then, one
applies a simultaneous ``$\beta$-reduction"
$(\lambda \vy.P_b) \vec{N} \longrightarrow
P_b [\vec{N}/\vy]$.

The {\em head variable} $x$ in an head normal form $x | \coa \langle N_1,\ldots, N_n \rangle$
plays the same role as a pointer in a strategy
does in Hyland-Ong games and
 an address (or locus) in
Girard's ludics.
On the other hand, a variable $x$ occurring in a bracket as in $N_0|\coa \langle N_1, \ldots,$ $N_{i-1}$, $x$, $ N_{i+1}, \ldots, N_n \rangle$
does not correspond to a pointer nor address. Rather, it corresponds to
an identity axiom (initial sequent) in sequent calculus,
and for this reason is called an \textbf{identity}.
If a negative design $N$ simply consists of a variable $x$,
then $N$ is itself an identity.

The positive design $\Omega$ denotes \emph{divergence}
(or \emph{partiality}) of the computation, in the sense   we will  make more precise
in the next subsection.
We also use $\Omega$ to encode partial sums.
Given a set $\alpha=\{a(\vx), b(\vy),c(\vz),\dots\}$
of negative actions with distinct names $\{a,b,c,\ldots\} \subseteq A$, we write
$\Sum_\alpha a(\vx).P_a$ to denote
the negative design $\Sum a(\vx).R_a$, where
$R_a=P_a$ if $a(\vx) \in \alpha$, and
$R_a=\Omega$ otherwise.
We also use an informal notation
$a(\vx).P_a + b(\vy). P_b + c(\vz).P_c + \cdots$ to denote $\Sum_\alpha a(\vx).P_a$.

So far, the syntax we are describing
is essentially the  same as the one introduced in \cite{Terui08}.
A novelty of this paper is the nondeterministic conjunction
operator $\Bigwedge$, which allows us
to build a positive design  $\Bigwedge \{S_i: i\in I\}$
from a set
 $\{S_i: i\in I\}$ of predesigns with $I$ an arbitrary index set.
Each $S_i$ is called a \textbf{conjunct} of $P$.
We write $\maltese$ ({\bf daimon})  for the  empty conjunction $\Bigwedge \emptyset$. This design plays an essential role in ludics,
since it is used to define the concept of orthogonality.
Although $\maltese$ is usually given
 as a primitive (see \eg \cite{DBLP:journals/mscs/Girard01,DBLP:journals/corr/abs-cs-0501039,Terui08}),
we have found it convenient and natural to
identify (or rather encode) $\maltese$ with the empty conjunction.
As we shall see, its  computational meaning   exactly corresponds  to the usual one: $\maltese$  marks the \emph{termination} of a computation.
Put in another way, our nondeterministic conjunction can be seen as
a generalization of the termination mark.


A design $D$ may
contain free and bound variables.
An occurrence of subterm $a(\vx).P_a$
\emph{binds} the free variables $\vx$ in $P_a$.
Variables which are not under the scope of the
binder $a(\vx)$ are \emph{free}. We denote
by $\mathsf{fv}(D)$ the set of free variables occurring in $D$.
As in $\lambda$-calculus,
we would like to identify two designs which are
\emph{$\alpha$-equivalent} \ie up to renaming of bound variables. But it is more subtle than usual, since
we also would like to identify, \eg  $\Bigwedge\{S, T\}$ with $\Bigwedge\{S\}$
whenever $S$ and $T$ are $\alpha$-equivalent.
To enforce these requirements simultaneously and hereditarily,
we define an equivalence relation by coinduction.

By {\em renaming} we mean a function
$\rho : \mathcal{V}\longrightarrow\mathcal{V}$.
We write $id$ for the identity renaming, and
$\rho[z/x]$ for
the renaming that agrees with $\rho$ except that
$\rho[z/x](x)=z$.
The set of renamings is denoted by $\mathcal{RN}$.

\begin{defi}[Design equivalence]\label{d-alpha}
A
binary relation $\cR \subseteq
(\mathcal{D}\times\mathcal{RN})^2$
is called a \textbf{design equivalence} if
for any $D, E, \rho, \tau$ such that $(D, \rho) \Rel (E, \tau)$,
one of the following holds:
\begin{enumerate}[(1)]
\item $D = \Omega = E$;
\item $D = \Bigwedge\{S_i : i\in I\}$,
$E = \Bigwedge\{T_j : j\in J\}$ and we have:
\begin{enumerate}[(i)]
\item for any $i\in I$ there is $j\in J$ such that
$(S_i,\rho) \Rel (T_j,\tau)$,
\item
for any $j\in J$ there is $i\in I$ such that
$(S_i,\rho) \Rel (T_j,\tau)$;
\end{enumerate}
\item $D = N_0|\Ov{a}\Bra{N_1, \dots, N_n}$,
$E = M_0|\Ov{a}\Bra{M_1, \dots, M_n}$
and $(N_k, \rho) \Rel (M_k, \tau)$
for every $0\leq k\leq n$;
\item $D= x$, $E= y$ and $\rho(x)=\tau(y)$;
\item $D= \Sum a(\vec{x}_a).P_a$,
$E= \Sum a(\vec{y}_a).Q_a$
and $(P_a, \rho[\vec{z}_a/\vec{x}_a])
\Rel
(Q_a, \rho[\vec{z}_a/\vec{y}_a])$
for every $a\in A$ and some vector
 $\vec{z}_a$ of fresh variables.
\end{enumerate}
We say that two designs $D$ and $E$ are \emph{equivalent}
if there is a design equivalence $\cR$ such that $(D, id) \Rel (E, id)$.
See \cite{Terui08} for further details.
\end{defi}

Henceforth we \emph{always} identify two designs $D$ and $E$,
and write $D=E$ by abuse of notation, if they are equivalent
in the above sense.
The following lemma is
a straightforward extension of
Lemma 2.6 of \cite{Terui08}.
It makes it easier to prove equivalence of two designs
(just as the ``bisimulation up-to'' technique in concurrency theory
makes it easier to prove bisimilarity of two processes).

\begin{lem}\label{l-equiv}
Let $\cR$ be a binary relation on designs such that
if $D\Rel E$ then one of the following holds:
\begin{enumerate}[\em(1)]
\item $D = \Omega = E$;
\item $D = \Bigwedge\{S_i : i\in I\}$,
$E = \Bigwedge\{T_j : j\in J\}$, and we have:
\begin{enumerate}[\em(i)]
\item for any $i\in I$ there is $j\in J$ such that
$S_i \Rel T_j$,
\item for any $j\in J$ there is $i\in I$ such that
$S_i \Rel T_j$;
\end{enumerate}
\item $D = N_0|\Ov{a}\Bra{N_1, \dots, N_n}$,
$E = M_0|\Ov{a}\Bra{M_1, \dots, M_n}$
and $N_k \Rel M_k$
for every $0\leq k\leq n$;
\item $D= x = E$;
\item $D=\Sum a(\vec{x}).P_a$,
$E=\Sum a(\vec{x}).Q_a$
and $P_a \Rel Q_a$ for every $a\in A$.
\end{enumerate}
If $D \Rel E$, then $D$ and $E$ are equivalent. \qed
\end{lem}

\noindent As a notational convention, a unary conjunction $\Bigwedge\{ S\}$ is simply written as $S$.
This allows us to treat a predesign as a positive design.
We also write:
 \begin{enumerate}[$\bullet$]
 \item
 $S \in P$ if $P$ is a conjunction and $S$ is a conjunct of $P$;
  \item  $P \leq Q$ if \emph{either} $P = \Omega$, \emph{or}
both $P$ and $Q$ are conjunctions and for all $S \in Q$, $S \in P$.
\end{enumerate}

\medskip\noindent Thus $P \leq Q$ indicates that $P$ has more conjuncts than $Q$ unless
$P = \Omega$.
We also extend the conjunction operator  to  positive designs and abstractions  as follows.

\begin{defi}[$\wedge$ operation] \label{conj binary}
\hfill
\begin{enumerate}[(1)]
\item As for positive designs, we set
$$
\begin{array}{rclrcl}
\Bigwedge\{S_i : i\in I\} \wedge \Bigwedge\{S_j : j\in J\} & := &
\Bigwedge\{S_k : k\in I\cup J\}, &
\Omega \wedge P & := & \Omega.
\end{array}
$$

\item As for abstractions,
\ie negative designs of the form $\Sum a(\vx).P_a$,
observe that since
 we are working up to renaming of bound variables,  it is no loss of generality to assume that  in any pair 
 $\Sum a(\vx).P_a$, $\Sum a(\vy).Q_a $, one has
$\vx = \vy$ for every $a \in A$.
We set:
$$\Sum a(\vx).P_a \wedge
\Sum a(\vx).Q_a  \; := \;  \Sum a(\vx).(P_a \wedge Q_a).$$
\end{enumerate}
\end{defi}

\medskip\noindent Observe the following:
\begin{enumerate}[$\bullet$]
\item The set of positive designs forms a semilattice with respect to $\leq$ and $\wedge$.
\item  $\Omega \leq P \leq \maltese$ for any positive  design $P$.
\end{enumerate}

\medskip\noindent The previous definition can be naturally generalized
to arbitrary sets as follows:

\begin{defi}[$\Bigwedge$ operation] \label{conj extension}\hfill
\begin{enumerate}[(1)]
\item Given a set $\bX$ of positive designs,
we define the positive design $\Bigwedge \bX$ as
follows:
\begin{enumerate}[$\bullet$]
\item If $\bX = \emptyset$, we set $\Bigwedge \bX := \maltese$.
\item  If   $\Omega \in \bX$, we set  $\Bigwedge \bX := \Omega$.
\item Otherwise, $\bX$ is a nonempty set of conjunctions
and we set:
$$ \Bigwedge \bX := \Bigwedge \{ S : S \in P \mbox{ for some } P \in \bX \}.$$
\end{enumerate}

\item Given a set  $\bX$  of  abstractions, we define the abstraction $\Bigwedge \bX$ as:
$$\Bigwedge \bX := \Sum a(\vx).\Bigwedge \{P_a : \Sum a(\vx).P_a \in \bX \}.$$
In particular, if $\bX = \emptyset$  then
$\Bigwedge \bX = \Sum a(\vx).\maltese$.
\end{enumerate}
\end{defi}

Notice  that $\Bigwedge \{D\} = D$, as long as $D$ ranges over  positive designs  or abstractions.


%
%
%

A design $D$ is said:
\begin{enumerate}[$\bullet$]
\item \textbf{total}, if $D \neq \Omega$;
\item \textbf{closed}, if $D$ has no occurrence of free variable;
\item \textbf{linear} (or \textbf{affine}, more precisely), if for any
subdesign of the form
$N_0|\coa \langle N_1,\ldots, N_n \rangle$, the sets
$\mathsf{fv}(N_0)$, \ldots, $\mathsf{fv}(N_n)$ are pairwise
disjoint;
\item\textbf{deterministic}, if in any occurrence
 of subdesign $\Bigwedge \{S_i : i \in I \}$, $I$ is either empty
(\ie we have $\maltese$) or  a singleton (\ie
we have a predesign).

\item \textbf{cut-free}, if it does not contain a cut as a subdesign;
\item \textbf{identity-free}, if it does not contain an identity
as  subdesign.
\end{enumerate}
%


We remark that the notion of design introduced in \cite{Terui08}
exactly corresponds in our terminology to that of  deterministic design.
Furthermore,  considering the specific  signature $\cG$  given below,
we can  also express in our setting Girard's original notion of design:

\begin{exa}[Girard's syntax]
Let us consider the signature $\cG=(\cP_{f}(\mathbb{N}),|\phantom{I}|)$
 where:
 \begin{enumerate}[$\bullet$]
 \item
 $\cP_{f}(\mathbb{N})$ consists of finite subsets
 of $\mathbb{N}$;
 \item  $|\phantom{I}| : \cP_{f}(\mathbb{N}) \longrightarrow \mathbb{N} $ is the function
 that maps a finite subset $ I \subseteq_f \mathbb{N}$ to its cardinality $|I| \in \mathbb{N}$.
 \end{enumerate}
 Girard's designs correspond to
total, linear, deterministic,  cut-free and identity-free designs
which have a finite number of free variables
over the signature $\cG$. See \cite{Terui08} for  more details.
\end{exa}

\subsection{Normalization} \label{normalization subsection}
Ludics is an interactive theory. This means that designs,
which subsume both proofs and models, interact together
via normalization, and types (behaviours) are defined by the
induced orthogonality relation (Section \ref{s-behaviour}).
 Several ways to normalize designs have been considered in the literature:
abstract machines \cite{DBLP:journals/mscs/Curien98,DBLP:conf/csl/Faggian02,DBLP:journals/corr/abs-0706-2544,Basaldella09},
abstract merging of orders \cite{DBLP:journals/mscs/Girard01,FP09}, and terms reduction \cite{DBLP:journals/corr/abs-0706-2544,Terui08}. Here we extend the last solution \cite{Terui08}.
As in untyped $\lambda$-calculus, normalization is  not necessarily terminating, but in our setting a
new difficulty  arises through the presence of
the operator $\Bigwedge$.

We define the normal forms in two steps, first giving a {\em nondeterministic}
reduction rule which finds head normal forms whenever possible,
and then expanding it corecursively. As usual,
let $D[\vec{N}/\vx]$ denote the design obtained by the simultaneous and capture-free
substitution of negative designs $\vec{N}=N_1,\ldots, N_n$ for $\vx=x_1,\ldots, x_n$
in $D$.



\begin{defi}[Reduction relation $\nondet$]
Given positive designs $P, Q$, we write $P \nondet Q$
 if
$\left(\Sum a(\vx).P_a\right) \; |\;  \cob \langle \vec{N} \rangle \in P$
and $Q = P_b[\vec{N}/\vx]$.
We denote by $\nondet^+$ the transitive closure,
and by $\nondet^\ast$ the reflexive transitive closure
of $\nondet$.
\end{defi}

Given two binary relations $\cR_1, \cR_2$ on designs, we write $\cR_1 \cR_2$ to denote the relation given by
their  composition \ie
$$D \ \cR_1 \cR_2 \ F \ \ \Longleftrightarrow \   \mbox{ there exists
a design }
E  \mbox { such that }  D \ \cR_1 \ E  \mbox{ and }  E \ \cR_2 \ F .$$
For instance,
we write $P\nondet^* \ni S$ if there exists $Q$ such
that $P\nondet^* Q$ and $Q\ni S$.

\begin{exas} We now give  examples of reductions and some remarks.
\begin{enumerate}[(1)]

    \item
$a(x).\maltese \;|\; \coa \langle K \rangle \ \nondet \  \maltese$.
\item
$b(x).\maltese \;|\;\coa \langle K \rangle  \ \nondet \  \Omega$ (recall that $b(x).\maltese$ stands for $b(x).\maltese + a(y).\Omega + c(\vz).\Omega + \cdots$ by our conventions on partial sums).
    \item Let $P =
a(x).\maltese \;|\; \coa \langle K \rangle \ \wedge \
b(x).\maltese \;|\;\coa \langle K \rangle$.
We have $P  \ \nondet \  \maltese$
and $P \ \nondet \  \Omega$.

\item For $N= a(x).x|\coa \langle x \rangle$, let  us consider $ P = N \;|\; \coa \langle N \rangle$.
We then  have an infinite reduction sequence
$$  P
\  \nondet \ P \  \nondet \ P
\  \nondet \ \cdots$$
since $ P= N \;|\; \coa \langle N \rangle   =
(a(x).x|\coa \langle x \rangle ) \;|\; \coa \langle N   \rangle$ and the latter design reduces to  $x|\coa \langle x \rangle  [N  /x]   =   N \;|\; \coa \langle N \rangle =P$.

\item Let   $P=a(y). (y|\cob \langle w \rangle \wedge z|\overline{c} \langle M \rangle ) \;|\; \coa \langle b(t).Q \rangle$.
 We have  the following reduction:
$$P \ \nondet \  (y|\cob \langle w \rangle \wedge z|\overline{c} \langle M \rangle ) [b(t).Q/y] \ = \ b(t).Q \;|\; \cob \langle w  \rangle \, \wedge \, z|\overline{c} \langle M[b(t).Q/y] \rangle.$$
We therefore have $P \nondet \ni  b(t).Q \;|\; \cob \langle w  \rangle$ and $P \nondet \ni z|\overline{c} \langle M[b(t).Q/y] \rangle$.
Since  $\ b(t).Q \;|\; \cob \langle w  \rangle$ is a cut,
  we have:
 $$ b(t).Q \;|\; \cob \langle w  \rangle \, \wedge \, z|\overline{c} \langle M[b(t).Q/y] \rangle \ \nondet \ Q[w/t].$$

\item The special designs $\maltese$ and $\Omega$ do not reduce to anything (as we will see, they are normal forms).

\item
By its definition, our reduction is not  ``closed under context" \ie
if
$P \nondet Q$ and $P$ (resp. $Q$) occurs as a subdesign
of $D$ (resp. $E$), nothing ensures that
$D \nondet E$. For instance a negative
design (or an head normal form)  having an occurrence of cut as
subdesign does not reduces to anything.
To  expand the reduction  ``under context"  we will use
Definition \ref{normal form}.

\end{enumerate}
\end{exas}


\noindent Notice that any {\em closed} positive design $P$
has
one of the following forms: $\maltese$, $\Omega$ and
$\Bigwedge\{S_i : i \in I\}$, where $S_i$ are cuts. The conjunction then
reduces to another closed positive design. Hence
any sequence of reductions starting from $P$ either  terminates with $\maltese$ or $\Omega$  or it diverges.
By {\em stipulating}
 that the normal form of $P$ in case of divergence is $\Omega$,
we obtain a dichotomy between $\maltese$ and $\Omega$:
the normal form of a closed positive design
is either $\maltese$ or $\Omega$.

This leads us to the following definition of normal form:


\begin{defi}[Normal form] \label{normal form}
The \textbf{normal form function} $\pl \ \pr : \cD \longrightarrow \cD$
is  defined by corecursion
as follows:
$$
\begin{array}{rcll}
\pl P \pr & = & \Omega  & \mbox{if  there is an infinite reduction}\\
& & & \mbox{sequence or a reduction sequence}\\
& & & \mbox{ending with $\Omega$ starting from $P$;}\\
 & = &
\Bigwedge \{ x|\coa \langle \pl \vec{N} \pr\rangle\; :\; P\nondet^* \ni
x|\coa \langle \vec{N}\rangle\}
& \mbox{otherwise;}\\
\pl \Sum a(\vx).P_a \pr & = & \Sum a(\vx). \pl P_a \pr;
&\\
\pl x \pr  & = & x.
\end{array}
$$
\end{defi}

\medskip\noindent We observe that when $P$ is a  closed positive design, we have
$\pl P \pr = \maltese$
precisely when \emph{all} reduction sequences from $P$ are finite and terminate with $\maltese$;
thus our nondeterminism is \emph{universal} rather
than existential.
This, however, does not mean that the set
$\{Q : P\nondet^* Q\}$ is finite; even when it is infinite,
it may happen that $\pl P \pr = \maltese$.

The following facts are easily observed:
\begin{lem}\label{l-facts} \hfill
\begin{enumerate}[\em(1)]
\item If $P\neq \Omega$, $P\leq Q$  and
$Q\nondet R$,
then $P\nondet R$.
\item
 $P\nondet Q$ implies $\pl P \pr \leq \pl Q \pr$. Furthermore,
if $P$ is a predesign, then $P \nondet Q$ implies
$\pl P \pr = \pl Q \pr$.
\item  $\pl \Bigwedge \bX \pr = \Bigwedge \{ \pl P \pr : P \in \bX \}$,
for any set $\bX$ of positive  designs.
\qed
\end{enumerate}
\end{lem}

\medskip\noindent Notice that the first statement means that
the composed relation $\leq \nondet$ is equivalent to
$\nondet$ as far as total designs are concerned.

\begin{exa}[Acceptance of finite trees]
In \cite{Terui08}, it is illustrated how
words and deterministic finite automata
are represented by (deterministic) designs in ludics.
We may extend the idea to trees and finite tree automata
in presence of nondeterminism. Rather than describing it in full detail,
we will only give an example which illustrates the power of nondeterminism
 to express (topdown) finite tree automata.

We consider the set of \emph{finite
trees} labelled with $a, b$ which are at most binary branching.
It is defined by the following grammar:
$$
t\; :: =\; \epsilon\; |\; a(t_1, t_2)\; |\; b(t_1, t_2).
$$
Here, $a(t_1, t_2)$ represents a tree with the root labelled by $a$
and with two subtrees $t_1, t_2$. In particular, $a(\epsilon, \epsilon)$
represents a leaf labelled by $a$. We simply write $a$ in this  case.

Suppose that the signature $\mathcal{A}$ contains a
unary name $\uparrow$, binary names $a, b$ and a nullary name $\epsilon$. We write $\downarrow$ for the positive action $\overline{\uparrow}$.
We abbreviate $\uparrowd(x). x|\coa\langle \vec{N}\rangle$ by
$\uparrowd\coa\langle \vec{N}\rangle$, so that we have
$$(\uparrowd\coa\langle \vec{N}\rangle)\; |\; \downarrowd\langle a(\vx).P\rangle
\
\nondet\
(a(\vx).P)\; |\; \coa\langle \vec{N}\rangle
\ \nondet\
P[\vec{N}/\vx].
$$

\medskip\noindent Each tree is then represented by a deterministic linear negative design
as follows:
\begin{eqnarray*}
\epsilon^\star & := & \uparrowd\overline{\epsilon},\\
a(t_1, t_2)^\star & := & \uparrowd\coa\langle t_1^\star, t_2^\star\rangle,\\
b(t_1, t_2)^\star & := & \uparrowd\cob\langle t_1^\star, t_2^\star\rangle.
\end{eqnarray*}

\medskip\noindent Now consider the positive design $Q= Q_0[x_0]$ defined by the following
equations:
\begin{eqnarray*}
Q_0[x] & := & x|\downarrowd\langle a(x,y). Q_1[x]\wedge Q_0[y]\; + \;
      b(x,y). Q_2[x]\wedge Q_2[y]\rangle, \\
Q_1[x] & := & x|\downarrowd\langle b(x,y). Q_2[x]\wedge Q_2[y]\rangle, \\
Q_2[x] & := & x|\downarrowd\langle \epsilon.\maltese\rangle.
\end{eqnarray*}
This design $Q$ works as an automata accepting all trees of
the form $a(b, a(b, \cdots a(b,b)\cdots))$.

Indeed, given $a(b,a(b,b))$,
it works nondeterministically as follows:
$$
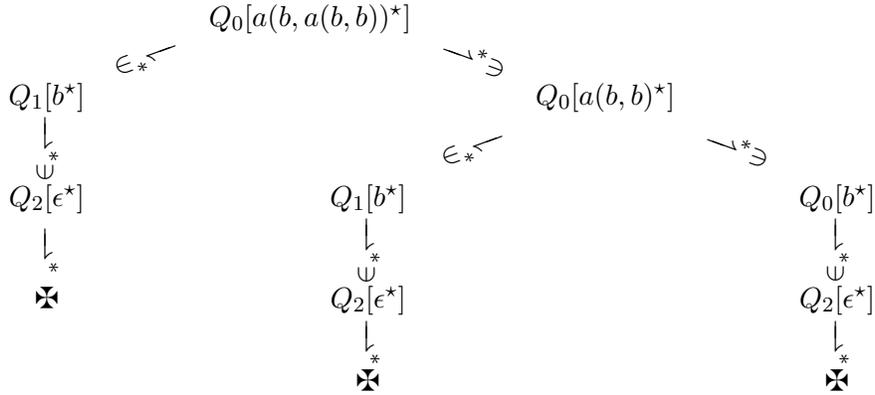
\begin{array}{cccccccc}
 & & Q_0 [a(b,a(b,b))^\star] \\
 & \rotatebox[origin=c]{200}{$\nondet^*\ni$} & & \rotatebox[origin=c]{340}{$\nondet^*\ni$} \\
 Q_1[b^\star] & & & & Q_0[a(b,b)^\star]\\
 \rotatebox[origin=c]{270}{$\nondet^*\ni$} & & & \rotatebox[origin=c]{200}{$\nondet^*\ni$} & & \rotatebox[origin=c]{340}{$\nondet^*\ni$} \\
Q_2[\epsilon^\star] & & \qquad\qquad Q_1[b^\star] & & & & Q_0[b^\star] \\
\rotatebox[origin=c]{270}{$\nondet^*$} & & \qquad\qquad \rotatebox[origin=c]{270}{$\nondet^*\ni$} & & & & \rotatebox[origin=c]{270}{$\nondet^*\ni$} \\
\maltese & & \qquad\qquad Q_2[\epsilon^\star] & & & & Q_2[\epsilon^\star] \\
 &  & \qquad\qquad \rotatebox[origin=c]{270}{$\nondet^*$} & & & & \rotatebox[origin=c]{270}{$\nondet^*$} \\
 &  & \qquad\qquad \maltese & & & & \maltese
\end{array}
$$
Hence we conclude $\Norm{Q_0 [a(b,a(b,b))^\star]} = \maltese$,
\ie $Q$ ``accepts'' the tree $a(b,a(b,b))$.
\end{exa}

\subsection{Associativity}

In this subsection, we prove one of the fundamental properties of
designs which we will need later:

\begin{thm}[Associativity]\label{t-assoc}
Let $D$ be a design
and $N_1,\ldots,N_n$ be negative designs.
We have:
$$ \pl D [N_1/y_1, \ldots, N_n/y_n] \pr = \pl\pl D \pr [\pl N_1 \pr/y_1, \ldots, \pl N_n \pr/y_n]\pr.$$
\qed
\end{thm}

Associativity corresponds to a weak form of the Church-Rosser property:
the normal form is the same even if we do not follow
the head reduction strategy. In this paper we are not concerned with
the full Church-Rosser property, and leave it as an open question.

The proof consists of several stages and it  can be skipped at first reading.


To prove associativity,
first notice that a simultaneous substitution
$D[N_1/y_1, \dots, N_n/y_n]$ can be turned into a
sequential one of the form
$D'[N_1/z_1]\cdots[N_n/z_n]$ by renaming $y_1, \dots, y_n$ by
fresh variables $z_1, \dots, z_n$ as follows:
$$D[N_1/y_1, \dots, N_n/y_n]
= D[z_1/y_1, \dots, z_n/y_n][N_1/z_1]\cdots[N_n/z_n].
$$
Moreover, we have:
$$\Norm{D}[\; \Norm{N_1}/y_1, \dots, \Norm{N_n}/y_n]
= \Norm{\; D[z_1/y_1, \dots, z_n/y_n]\; }[\; \Norm{N_1}/z_1]\cdots
[\; \Norm{N_n}/z_n].
$$
This allows us to work with sequential substitutions rather than simultaneous
ones.

We define a binary relation $\gg$ on designs by:
\begin{enumerate}[$\bullet$]
\item $D\gg E$ if $D=D_0[N_1/y_1] \cdots [N_n/y_n]$ and
$E= \Norm{D_0}[\Norm{N_1}/y_1]\cdots[\Norm{N_n}/y_n]$
for some $D_0, N_1, \dots, N_n$ such that $y_i\not\in \mathsf{fv}(N_j)$
for $1\leq i\leq j\leq n$.
\end{enumerate}

\begin{lem}\label{l-a1}
Suppose that
$P=P_0[N_1/y_1]\cdots[N_n/y_n]$ and
$Q=\Norm{P_0}[\Norm{N_1}/y_1]\cdots[\Norm{N_n}/y_n]$ so that $P \gg Q$.
When $P \nondet P'$, two cases can be distinguished.
\begin{enumerate}[\em(1)]
\item If $P_0 \nondet P_1$ and $P' = P_1 [N_1/y_1]\cdots[N_n/y_n]$,
then there exists $Q'$ such that $Q\leq Q'$ and $P' \gg Q'$.
\item Otherwise, there exists
$Q'$ such that $Q\nondet Q'$ and $P' \gg Q'$.
\end{enumerate}
\end{lem}

\proof \hfill
\begin{enumerate}[(1)]
\item  By Lemma \ref{l-facts} (2), we have
$\Norm{P_0}\leq\Norm{P_1}$ that implies
$Q \leq \Norm{P_1}[\Norm{N_1}/y_1]\cdots[\Norm{N_n}/y_n]$.
Hence by letting $Q' =
\Norm{P_1}[\Norm{N_1}/y_1]\cdots[\Norm{N_n}/y_n]$, we have
$Q\leq Q'$ and $P' \gg Q'$.

\item
If (1) is not the case,
a cut must be created by substitution of some $N_j$ for
a head variable of $P_0$. Hence
$P_0$ must contain a head normal form
$y_j |\Ov{a}\Bra{\vec{M}}$ as conjunct for
some $1\leq j \leq n$ and
$N_j = \Sum a(\vec{x}).R_a$, so that
$P$ contains a cut
$$y_j |\Ov{a}\Bra{\vec{M}}[N_1/y_1]\cdots[N_n/y_n]
= N_j |\Ov{a}\Bra{\vec{M}}[N_1/y_1]\cdots[N_n/y_n]
$$
(the equality due to $y_i\not\in \mathsf{fv}(N_j)$ for $1\leq i\leq j$) and
$P \nondet
R_a[\vec{M}/\vec{x}][N_1/y_1]\cdots[N_n/y_n] $ $= P'$.
In this case, $\Norm{P_0}$ contains $y_j |\Ov{a}\Bra{\Norm{\vec{M}}}$
so that $Q$ contains
$$
y_j |\Ov{a}\Bra{\Norm{\vec{M}}}
[\Norm{N_1}/y_1]\cdots[\Norm{N_n}/y_n]
  =
\Norm{N_j} |\Ov{a}\Bra{\Norm{\vec{M}}}
[\Norm{N_1}/y_1]\cdots[\Norm{N_n}/y_n].$$
Since $\Norm{N_j} = \Sum a(\vec{x}).\Norm{R_a}$,  we have $Q \nondet
\Norm{R_a}[\vec{\Norm{M}}/\vec{x}][
\Norm{N_1}/y_1] \cdots[\Norm{N_n}/y_n]$.
Let $Q'$ be the latter design.
Since the simultaneous substitutions
$[\vec{M}/\vec{x}]$
and
$[\vec{\Norm{M}}/\vec{x}]$
can be made sequential,
we have $P'\gg Q'$.
\qed
\end{enumerate}

\begin{lem}\label{l-a2}
If $P \gg Q$ and
$Q\nondet Q'$, then there exists some $P'$
such that $P'\gg Q'$ and $P \nondet^+ P'$.
\end{lem}

\proof
Suppose that
$P=P_0[N_1/y_1]\cdots[N_n/y_n]$,
$Q=\Norm{P_0}[\Norm{N_1}/y_1]\cdots[\Norm{N_n}/y_n]$
and $Q\nondet Q'$.
Then $\Norm{P_0}$ must contain
$y_j |\Ov{a}\Bra{\Norm{\vec{M}}}$
for some $\vec{M}$ and $1\leq j\leq n$.
Thus, $P_0 \nondet^*\ni y_j |\Ov{a}\Bra{\vec{M}}$.
Suppose also
$N_j = \Sum a(\vec{x}).R_a$ so that
$\Norm{N_j} = \Sum a(\vec{x}).\Norm{R_a}$.

Now the situation is as follows: $Q$ contains
$$y_j |\Ov{a}\Bra{\Norm{\vec{M}}}
[\Norm{N_1}/y_1]\cdots[\Norm{N_n}/y_n]
=
\Norm{N_j} |\Ov{a}\Bra{\Norm{\vec{M}}}
[\Norm{N_1}/y_1]\cdots[\Norm{N_n}/y_n],$$
so we have
$$Q\nondet
\Norm{R_a}[\vec{\Norm{M}}/\vec{x}]
[\Norm{N_1}/y_1] \cdots [\Norm{N_n}/y_n] = Q'.$$
On the other hand,
$$ \begin{array}{rcrl}
P & \nondet^* \ni &
y_j |\Ov{a}\Bra{\vec{M}}[N_1/y_1] \cdots [N_n/y_n] & \\
& = & N_j |\Ov{a}\Bra{\vec{M}}[N_1/y_1] \cdots [N_n/y_n] & \\
& \nondet &
R_a[\vec{M}/\vec{x}_a] [N_1/y_1] \cdots [N_n/y_n] & = P',
\end{array}$$
which implies $P\nondet^+ P'$.
Since the simultaneous substitutions
$[\vec{M}/\vec{x}]$
and
$[\vec{\Norm{M}}/\vec{x}]$
can be made sequential,
we have $P'\gg Q'$.
\qed

\begin{lem}\label{l-a3}
Suppose that $P \gg Q$. Then
$\Norm{P}=\Omega$ if and only if $\Norm{Q} = \Omega$.
\end{lem}

\proof \hfill
\begin{enumerate}[$\bullet$]
\item
For the `if' direction, we distinguish two cases.
\begin{enumerate}[$-$]
\item If there is an infinite reduction sequence from $Q$,
then there is also an infinite sequence from $P$ by Lemma \ref{l-a2}.
\item If $Q \nondet^* \Omega$, then there is $P'$ such that
$P \nondet^* P'$ and $P' \gg \Omega$.
Namely, $P'$ can be written as $P_0[N_1/y_1]\cdots[N_n/y_n]$ and
$\Omega =\Norm{P_0}[\Norm{N_1}/y_1]\cdots[\Norm{N_n}/y_n]$.
The latter means that $\Norm{P_0} = \Omega$, which implies
$\Norm{P'} = \Omega$. From this and
$P \nondet^* P'$, we conclude $\Norm{P} = \Omega$.
\end{enumerate}
\item For the `only-if' direction, if $P \nondet^* \Omega$, we easily obtain
$\Norm{Q} =\Omega$. Otherwise,
there is an infinite reduction sequence
$P= P^0 \nondet P^1 \nondet P^2 \nondet \cdots$.
Suppose that $P = P^0 = P_0[N_1/y_1]\cdots[N_n/y_n]$ and
$Q =\Norm{P_0}[\Norm{N_1}/y_1]\cdots[\Norm{N_n}/y_n]$.
Our purpose is to build either a finite reduction sequence
$Q\nondet^* \Omega$ or
an infinite reduction sequence
$Q = Q^0 \nondet Q^1 \nondet Q^2 \nondet \cdots$.
Two cases arise:
\begin{enumerate}[$-$]
\item The reductions take place inside $P_0$ and
independently of $N_1, \dots, N_n$.
Namely, there is an infinite reduction sequence
$P_0 \nondet P_1 \nondet P_2 \nondet \cdots$
such that $P^i = P_i[N_1/y_1]\cdots[N_n/y_n]$ for every $i\geq 0$.
Then $\Norm{P_0}= \Omega$, which implies $Q=\Omega$.
So we have $\Norm{Q}=\Omega$.
\item Otherwise,
there is at most a finite sequence
$P_0 \nondet P_1 \nondet \cdots \nondet P_m$ such that
$P^i = P_i[N_1/y_1]\cdots[N_n/y_n]$ for $0\leq i\leq m$ and $P_m$ contains a head
normal form that is responsible for the reduction $P^m \nondet P^{m+1}$.
By repeatedly applying Lemma \ref{l-a1} (1),
we obtain $Q'$ such that $Q\leq Q'$ and
$P^m \gg Q'$.
Since $P^m \nondet P^{m+1}$, there exists $Q^{1}$ such that
$Q' \nondet Q^{1}$ and $P^{m+1} \gg Q^{1}$
by Lemma \ref{l-a1} (2). Hence
by Lemma \ref{l-facts} (1), we obtain $Q= Q^0 \nondet Q^{1}$.
\end{enumerate}
In the former case, we are already done.
In the latter case,
we still have an infinite reduction sequence
$P^{m+1} \nondet P^{m+2} \nondet \cdots$ and
$P^{m+1} \gg Q^{1}$. Hence we may repeat
the same argument to prolong the reduction sequence
$Q^0\nondet Q^1$.
Hence we eventually obtain $\Norm{Q}=\Omega$.
\qed
\end{enumerate}

\begin{lem}\label{l-a4}
Suppose that $P \gg Q$.
\begin{enumerate}[$\bullet$] \item If $P \nondet^* \ni
x|\Ov{a}\Bra{M_1, \dots, M_m}$, then there exist $L_1, \dots, L_m$
such that
$Q\nondet^*\ni x|\Ov{a}\Bra{L_1, $ $\dots, L_m}$ and
$M_1 \gg L_1$, \dots, $M_m \gg L_m$.
\item
Conversely,
if $Q \nondet^* \ni
x|\Ov{a}\Bra{L_1, \dots, L_m}$,
then there exist $M_1, \dots, M_m$
such that
$P\nondet^*\ni x|\Ov{a}\Bra{M_1, \dots, M_m}$ and
$M_1 \gg L_1$, \dots, $M_m \gg L_m$.
\end{enumerate}
\end{lem}

\proof
Suppose that $P \nondet^* P'
\ni
x|\Ov{a}\Bra{M_1, \dots, M_m}$.
By Lemmas \ref{l-a1} and \ref{l-facts} (1)
(which states that the composed relation
$\leq\nondet$ is identical with $\nondet$),
there is  $Q'$  such that $Q\nondet^* \leq Q'$ and
$P' \gg Q'$. Since
$P'\ni
x|\Ov{a}\Bra{\vec{M}}$, we may write
$$
x|\Ov{a}\Bra{\vec{M}}=
x|\Ov{a}\Bra{\vec{K}}[N_1/y_1]\cdots [N_n/y_n]
=
x|\Ov{a}\Bra{\vec{K}[N_1/y_1]\cdots [N_n/y_n]}
$$
for some
$\vec{K} = K_1, \dots, K_m$, where $x\not\in \{y_1, \dots, y_n\}$, and
$Q'$ contains
$$\Norm{x|\Ov{a}\Bra{\vec{K}}}
[\Norm{N_1}/y_1]\cdots [\Norm{N_n}/y_n] =
x|\Ov{a}\Bra{\vec{\Norm{K}}[\Norm{N_1}/y_1]\cdots [\Norm{N_n}/y_n]}.$$
Hence by letting $L_i = \Norm{K_i}[\Norm{N_1}/y_1]\cdots [\Norm{N_n}/y_n]$
we obtain $M_i \gg L_i$ for every $1\leq i\leq m$. Since
$Q\nondet^*\leq Q' \ni x|\Ov{a}\Bra{\vec{L}}$, namely
$Q\nondet^* \ni x|\Ov{a}\Bra{\vec{L}}$,
the claim holds.

Conversely, suppose that
$Q \nondet^* Q'
\ni
x|\Ov{a}\Bra{L_1, \dots, L_m}$.
By Lemma \ref{l-a2},
there is  $P'$  such that $P\nondet^* P'$ and
$P' \gg Q'$. The rest is similar to the above.
\qed

\begin{lem}\label{l-a5}
If $M\gg N$, then either $M=y=N$ for a variable $y$, or
$M=\Sum a(\vec{x}_a).P_a$,
$N=\Sum a(\vec{x}_a).Q_a$ and
$P_a \gg
Q_a$ for every $a\in A$.
\end{lem}

\proof
Immediate.
\qed

The following lemma completes the proof of Theorem \ref{t-assoc}.

\begin{lem}
If $D_0\gg E_0$, then $\Norm{D_0} = \Norm{E_0}$.
\end{lem}

\proof
Define a binary relation $\cR$ on designs as follows:
\begin{enumerate}[$\bullet$]
\item For positive (resp.\ negative) designs $D, E$, we have
$D\Rel E$ if $D= \Norm{D_0}$, $E=\Norm{E_0}$, and $D_0 \gg E_0$ for
some $D_0$ and $E_0$.
\item For predesigns $S, T$, we have
$S\Rel T$ if $S=x|\Ov{a}\Bra{\Norm{M_1}, \dots, \Norm{M_m}}$,
$T=x|\Ov{a}\Bra{\Norm{L_1}, \dots, $ $  \Norm{L_m}}$, and
$M_i \gg L_i$ for every $1\leq i \leq m$.
\end{enumerate}

\medskip\noindent We now verify that this $\cR$ satisfies the conditions of
Lemma \ref{l-equiv}.

First, let $P, Q$ be positive designs such that $P \Rel Q$, \ie
$P= \Norm{P_0}$, $Q=\Norm{Q_0}$, and $P_0 \gg Q_0$ for
some $P_0$ and $Q_0$.
\begin{enumerate}[$\bullet$]
\item If $\Norm{P_0}=\Omega$, then
$\Norm{Q_0}= \Omega$ too by Lemma \ref{l-a3}.
Hence (1) holds.

\item If $\Norm{P_0}$ is a conjunction, then $\Norm{Q_0}$ is not $\Omega$
by Lemma \ref{l-a3},
so is a conjunction.
If $\Norm{P_0}$ contains
$x|\Ov{a}\Bra{\Norm{M_1}, \dots, \Norm{M_m}}$,
then
$P_0 \nondet^*\ni x|\Ov{a}\Bra{\vec{M}}$. Since $P_0 \gg Q_0$,
Lemma \ref{l-a4} yields
$Q_0 \nondet^*\ni x|\Ov{a}\Bra{\vec{L}}$ for some
$\vec{L}= L_1, \dots, L_m$.
Namely, $\Norm{Q_0}$ contains
$x|\Ov{a}\Bra{\Norm{\vec{L}}}$.
Moreover, we have
$M_i \gg L_i$
 for every $1\leq i \leq m$.
Similarly, one can show that
if $\Norm{Q_0}$ contains
$x|\Ov{a}\Bra{\Norm{\vec{L}}}$,
then $\Norm{P_0}$ contains
$x|\Ov{a}\Bra{\Norm{\vec{M}}}$
and $M_i \gg L_i$
 for every $1\leq i \leq m$.
Hence (2) holds.
\end{enumerate}
Let $S, T$ be predesigns such that $S \Rel T$, i.e.,
$S=x|\Ov{a}\Bra{\Norm{M_1}, \dots, \Norm{M_m}}$,
$T=x|\Ov{a}\Bra{\Norm{L_1}, \dots, $ $ \Norm{L_m}}$, and
$M_i \gg L_i$ for every $1\leq i \leq m$.
It immediately follows that
$\Norm{M_i} \Rel \Norm{L_i}$ for every $1\leq i \leq m$.
Also, $x \Rel x$. Hence (3) holds.

Finally, let $N, M$ be negative designs such that
$N= \Norm{N_0}$, $M=\Norm{M_0}$, and $N_0 \gg M_0$ for
some $N_0$ and $M_0$.
\begin{enumerate}[$\bullet$]
\item If $N=x$, then $N_0 = M_0 = M = x$.
Hence (4) holds.

\item Otherwise, $N$ must be of the form $\Sum a(\vec{x}_a).
\Norm{P_a}$ and $N_0 =
\Sum a(\vec{x}_a).P_a$. Since $N_0 \gg M_0$,
$M_0$ is of the form $\Sum a(\vec{x}_a).Q_a$ and $P_a \gg Q_a$
for every $a\in A$ by Lemma \ref{l-a5}. So
$M = \Norm{M_0} = \Sum a(\vec{x}_a).\Norm{Q_a}$ and $\Norm{P_a} \Rel \Norm{Q_a}$.
Hence (5) holds.
\end{enumerate}
Therefore, if $D_0 \gg E_0$,
we have $\Norm{D_0} = \Norm{E_0}$ by Lemma \ref{l-equiv}.
\qed

\section{Behaviours}\label{s-behaviour}

\noindent This section is concerned with the type structure of
ludics. We describe orthogonality and behaviours in \ref{ortho-s},
logical connectives in \ref{ss-logical} and finally explain
(the failure of)
internal completeness of logical connectives
in \ref{ss-internal}.

\subsection{Orthogonality} \label{ortho-s}

In the rest of this paper,
we mainly restrict ourselves to a special subclass
of designs:  we only consider  designs which are  \emph{total},
\emph{cut-free},
 and \emph{identity-free}. Generalizing the terminology in \cite{Terui08},
we call them {\em standard designs}. In other words:

\begin{defi}[Standard design]\label{d-standard}
 A design $D$  is said {\bf standard} if it satisfies the following two conditions:
\begin{enumerate}[(i)]
\item{\em Cut-freeness and identity-freeness}:
  $D$ can be coinductively generated by
the following restricted version of the
grammar given in Definition \ref{designs}:
$$
\begin{array}{rcl}
P  & ::=    & \Omega \ \big| \  \Bigwedge \{S_i : i\in I\},  \\
S & ::= &
  x |\coa \langle N_1,\ldots,N_n \rangle,  \\
 N & ::= &  \Sum a(\vx).P_a. \\
 \end{array}
$$
\item{\em Totality}:  $D \neq \Omega$.
\end{enumerate}
\end{defi}

\medskip\noindent The totality condition is due to
the original work \cite{DBLP:journals/mscs/Girard01}.
It has a pleasant consequence
that behaviours (see below) are never empty.
We also remark that the lack of
identities can be somehow compensated by considering
their infinitary $\eta$ expansions,
called  \emph{faxes} in \cite{DBLP:journals/mscs/Girard01}.
In our setting, the infinitary $\eta$ expansion of an identity
 $x$ is expressed by
the negative standard design $\eta(x)$  defined by the equation:
$$
\eta(x) = \Sum a(y_1,\ldots,y_n).x|\coa \langle \eta(y_1),\ldots,\eta(y_n) \rangle.$$
We refer to
\cite{Terui08} for more details.


We are now ready to define  orthogonality  and behaviours.

\begin{defi}[Orthogonality]
A positive  design $P$ is said
\textbf{atomic} if it is standard and $\mathsf{fv}(P) \subseteq \{x_0\}$
for a certain fixed variable $x_0$.\footnote{
The variable $x_0$ here
plays the same role as the empty
address ``$\langle \rangle$"  does in \cite{DBLP:journals/mscs/Girard01}
: $x_0$ may be thought of as a fixed and predetermined
``location."}

A negative design $N$ is said \textbf{atomic} if  it is standard and $\mathsf{fv}(N)=\emptyset$.

Two atomic designs $P,N$ of opposite polarities
are said \textbf{orthogonal} and written $P \bot N$ (or equivalently $N \bot P$) when
$\pl P[N/x_0] \pr = \maltese$.

If $\bX$ is a set of atomic designs of the same polarity, then
its \textbf{orthogonal set}, denoted by $\bX\b$, is defined by
$ \bX \b := \{ E : \forall D \in \bX, \ D \bot E \}$.
\end{defi}

\medskip\noindent The meaning of $\Bigwedge$ and the associated partial order $\leq$
can be clarified in terms of orthogonality.
For atomic designs $D, E$ of the same polarity, define
$D \preceq E$ if and only if $\{D\}^\bot \subseteq \{E\}^\bot$.
$D \preceq E$ means that $E$ has more chances of convergence than $D$
when interacting with other atomic designs.
The following
is easy to observe.
\begin{prop} \label{conj and ort}~
\begin{enumerate}[\em(1)]
\item $\preceq$ is a preorder.
\item $P \leq Q$ implies $P \preceq Q$ for any pair of atomic positive designs $P, Q$.
\item  Let $\bX$ and $\bY$ be sets of
atomic designs of the same polarity.
Then $\bX \subseteq \bY$ implies $\Bigwedge \bY \preceq
\Bigwedge \bX$.
\qed
\end{enumerate}
\end{prop}


In particular,
 $\Omega \preceq
P \preceq \maltese$ for any atomic
positive design $P$.\footnote{ Here we are
tentatively considering the nontotal design $\Omega$,
which
does not  officially belong to the universe of atomic  designs.}
This justifies our identification of $\maltese$ with the
empty conjunction $\bigwedge \emptyset$.


%

\begin{rem}\label{r-separation} Designs in
\cite{DBLP:journals/mscs/Girard01} satisfy the \emph{separation property}:
for any designs $D, E$ of the same polarity,
we have $D = E$ if and only if $\{D\}^\bot = \{E\}^\bot$.
But when   the constraint of
linearity is removed,
this property  no more holds, as observed in \cite{MauTh} (see also \cite{DBLP:journals/corr/abs-0706-2544}).

In our setting, separation  does not hold, even  when
$D$ and $E$ are
deterministic  (atomic)  designs.
For instance, consider the following two designs \cite{MauTh}:
\begin{eqnarray*}
P & := &
x_0|\mathord{\downarrow} \langle \mathord{\uparrowd}(y).\maltese \rangle,\\
Q & := & x_0|\mathord{\downarrow} \langle \mathord{\uparrowd}(y).P \rangle
\ = \
x_0|\mathord{\downarrow} \langle \; \mathord{\uparrowd}(y).\,
x_0|\mathord{\downarrow} \langle  \mathord{\uparrowd}(y).\maltese \rangle
\; \rangle.
\end{eqnarray*}
It is easy to see that in our setting $P \bot N$ holds if and only if
$N$ has an additive component
 of the form  $\uparrowd(z).\Bigwedge \{ z|\downarrowd \langle M_i \rangle : i \in I \}$ for   arbitrary index
set $I$ and arbitrary standard negative designs $M_i$ with $\mathsf{fv}(M_i) \subseteq\{z\}$.

The same holds for $Q$, as can be observed from the following reduction
sequence (for readability, we only consider the case in which $N$ has
a component of the form $\uparrowd(z). z|\downarrowd \langle M \rangle$, the general case
easily follows):
\begin{eqnarray*}
Q[N/x_0]   & = &
N \;|\; \downarrowd\langle\uparrowd(y). P [N/x_0]\rangle    \\
 & \nondet & \left( \uparrowd(y). P [N/x_0]\right) \;|\; \downarrowd\langle M[\uparrowd(y). P [N/x_0]/z]\rangle   \\
 & \nondet &  P[N/x_0]\ \nondet^\ast \ \maltese.
\end{eqnarray*}
(If $N$ does not have
a component of the form discussed above,
we have $Q[N/x_0] \nondet^* \Omega$.)

We therefore conclude $\{P\}^\bot = \{Q\}^\bot$, even though
$P\neq Q$. 
\end{rem}

\medskip\noindent Although possible, we do not define orthogonality for nonatomic
designs. Accordingly, we only consider \emph{atomic} behaviours
which consist of atomic designs.

\begin{defi}[Behaviour] \label{d-beh}
A \textbf{behaviour} $\bX$ is a set of atomic standard designs of the same
polarity
such that $\bX\b\b=\bX$.
\end{defi}
A behaviour is positive or negative according to the polarity of its designs.
We denote positive  behaviours by $\bP, \bQ, \bR,\dots$
and negative behaviours by $\bN, \bM, \bK \dots$.

Orthogonality satisfies
the following standard properties:

\begin{prop}  \label{closure operator}
Let $\bX,\bY$ be sets of atomic designs
of the same polarity. We have:
\begin{enumerate}[\em(1)]
\item $\bX \subseteq \bX\b\b$.
\item $\bX \subseteq \bY  \Longrightarrow \bY\b \subseteq \bX\b$.
\item $\bX \subseteq \bY\b\b \Longrightarrow \bX\b\b \subseteq \bY\b\b$.
\item $\bX\b = \bX\b\b\b$. In particular, any orthogonal set is a behaviour.
\item $(\bX \cup \bY)\b = \bX\b \cap \bY\b$. In particular, the intersection
of two behaviours is a behaviour. \qed
\end{enumerate}
\end{prop}

We also observe that $D\preceq E$ and $D\in \bX$ implies $E\in \bX$
when $\bX$ is a behaviour.

Among all positive
(resp.\ negative) behaviours, there exist the least and the greatest behaviours
with respect to set inclusion:
$$
\begin{array}{rclclcrclcl}
\zero^+ \!\!& \!\!:=\!\! & \{\maltese\}\b\b & \!\!\!\!=\!\!\!\! & \{\maltese\}, & \qquad   & \top^- \!\!& \!\!:=\!\! &\left(\zero^+ \right)\b & \!\!\!\!=\!\!\!\! & \left\{ \mbox{ atomic negative designs }  \right\} ,  \\
\zero^- \!\!& \!\!:=\!\! & \{\maltese^-\}\b\b & \!\!\!\!=\!\!\!\! & \{\maltese^-\}, & \qquad  & \top^+ \!\!& \!\!:=\!\! & \left(\zero^- \right)\b & \!\!\!\!=\!\!\!\! & \left\{ \mbox{ atomic positive designs } \right\},
 \\
\end{array}
$$
where
$\maltese^- := \Sum a(\vec{x}).\maltese$ plays the role of
 the design called \emph{negative daimon} in \cite{DBLP:journals/mscs/Girard01}.
Notice that behaviours are  always nonempty due to the
totality condition: any positive (resp.\ negative) behaviour
 contains $\maltese$ (resp.\ $\maltese^-$).

Now that we have given  behaviours, we can   define \emph{contexts of behaviours}
and  then the   \emph{semantical entailment}
$\models$ in order to  relate designs to contexts
of behaviours.
These constructs  play the role of typing environments in type systems.
They correspond to \emph{sequents of behaviours}, in the terminology
of \cite{DBLP:journals/mscs/Girard01}.

\begin{defi}[Contexts of behaviours and semantical entailment $\models$]
\hfill
\begin{enumerate}[(a)]\label{d-context}
\item A \textbf{positive context} $\bGamma$ is of the form
$x_1: \bP_1, \dots, x_n:\bP_n$,
where $x_1, \dots, x_n$ are distinct variables and
$\bP_1, \dots, \bP_n$ are (atomic) positive behaviours.
We denote by $\mathsf{fv}(\bGamma)$ the set $\{x_1, \dots, x_n\}$.

A \textbf{negative context} $\bGamma, \bN$ is a positive context $\bGamma$
enriched with an (atomic) negative behaviour $\bN$, to which no variable is
associated.
\item
The \textbf{semantical entailment} is the binary relation  $\models$ between designs and contexts of behaviours of the same polarity defined as follows:

%

\noindent  $P\models x_1: \bP_1, \dots, x_n:\bP_n$ if and only if:
\begin{enumerate}[$\bullet$]
\item $P$ is standard;
 \item $\mathsf{fv}(P)\subseteq \{x_1, \dots, x_n\}$;
\item$\pl P[K_1/x_1, \dots, K_n/x_n]\pr = \maltese$
for any $K_1 \in \bP_1^\bot$, \ldots, $K_n\in \bP_n^\bot$.
\end{enumerate}

\noindent $N\models x_1: \bP_1, \dots, x_n:\bP_n, \bN$ if and only if:
\begin{enumerate}[$\bullet$]
\item $N$ is standard;
\item
$\mathsf{fv}(N)\subseteq \{x_1, \dots, x_n\}$;
\item $\pl Q[N[K_1/x_1, \dots,  K_n/x_n]/x_0]\pr = \maltese$
for any $K_1 \in \bP_1^\bot$, \ldots, $K_n\in \bP_n^\bot$, $Q\in \bN^\bot$.
\end{enumerate}

\end{enumerate}
\end{defi}

\medskip\noindent Clearly,
$N\models \bN$ if and only if $N\in \bN$,
 and $P \models y : \bP$ if and only if $P[x_0/y]\in \bP$.
Furthermore,
associativity (Theorem \ref{t-assoc}) implies the following quite useful
principle:

\begin{lem}[Closure principle]\label{l-closure} \hfill
\begin{enumerate}[\em(1)]
\item $P \models  \bGamma, z : \bP$ if and only if  $\pl P[M/z] \pr
 \models \bGamma$ for
any $M \in \bP\b$;
\item $N  \models \bGamma, \bN $ if and only if $\pl Q[N/x_0] \pr \models  \bGamma$
for any $Q \in \bN\b$;
\item $N \models  \bGamma, z : \bP, \bN$ if and only if  $\pl N[M/z] \pr
 \models \bGamma,\bN$ for
any $M \in \bP\b$.
\end{enumerate}
\end{lem}

\proof \hfill
\begin{enumerate}[(1)]
\item
Let $P$ be a standard design with
$\mathsf{fv}(P) \subseteq
\{x_1, \dots, x_n,z\}$ and
$\bGamma$ a context $x_1 : \bP_1,\ldots, x_n : \bP_n$.

First, we claim that
$\pl P[M/z] \pr$ is a standard design when
$P \models  \bGamma, z : \bP$ and $M \in \bP\b$.
Indeed, it is obviously cut-free. It is
also identity-free
because so are $P,M$ and neither substitution
 $P[M/z]$ nor normalization $\pl P[M/z] \pr$ introduces
identities. Totality will be shown below.
We also note that
$\mathsf{fv}(\pl P[M/z] \pr) \subseteq \{x_1, \dots, x_n\}$,
since $M$ is an atomic negative design that is always closed.

Next, we observe that $\pl P[\vec{K}/\vx,M/z]\pr =
\pl\; \pl P[M/z]\pr\;  [\vec{K}/\vx] \pr$
for any list
$\vec{K} = K_1, \dots, K_n$
of standard negative designs.
Indeed, notice that
 $P[\vec{K}/\vx,M/z] = P[M/z]  [\vec{K}/\vx]$
since $M$ is closed, and
$\Norm{K_i} = K_i$ since $K_i$ is cut-free. Hence by associativity,
we obtain:
$$
\pl P[\vec{K}/\vx,M/z]\pr =
 \pl P[M/z]  [\vec{K}/\vx] \pr
=
\pl\; \pl P[M/z] \pr\;  [ \pl \vec{K} \pr /\vx] \pr
=
\pl\; \pl P[M/z] \pr\;  [  \vec{K} /\vx] \pr. 
$$
In particular,
$\pl P[\vec{K}/\vx,M/z]\pr = \maltese$ implies
the totality of  $\pl P[M/z] \pr$.

We are now ready to prove the first claim.
Writing $\vec{K}\in \bGamma\b$ for
$ K_1 \in \bP_1^\bot$, \ldots, $K_n \in \bP_n^\bot$, we have:
\begin{eqnarray*}
P \models \bGamma, z : \bP & \Longleftrightarrow &
\pl P[\vec{K}/\vx,M/z]\pr = \maltese \mbox{ for every $\vec{K}\in \bGamma\b$
and $M\in \bP^\bot$}, \\
& \Longleftrightarrow &
\pl\; \pl P[M/z] \pr\;  [  \vec{K} /\vx] \pr
= \maltese \mbox{ for every $\vec{K}\in \bGamma\b$
and $M\in \bP^\bot$}, \\
& \Longleftrightarrow &
\pl P[M/z] \pr \models \bGamma
\mbox{ for every $M\in \bP^\bot$.}
\end{eqnarray*}

\item and (3) are proven in a similar way.
We just mention that the crucial  equalities
\begin{eqnarray*}
\pl Q[N[\vec{ K }/\vx]/x_0] \pr & = &
\pl \; \pl Q[N/x_0] \pr  \; [\vec{ K }/\vx] \pr,\\
\pl Q[N[\vec{K}/\vx,M/z]/x_0]\pr & = &
\pl  Q [ \; \pl  N[M/z]\pr \; [\vec{K}/\vx]/x_0]\pr,
\end{eqnarray*}
which are needed to show  (2) and (3) respectively,
can be straightforwardly  derived  from associativity.
\qed
\end{enumerate}

\subsection{Logical connectives}\label{ss-logical}
We next describe how to build behaviours by means of {\em logical
connectives} in ludics.

\begin{defi}[Logical connectives]\label{d-lc}
An \textbf{$n$-ary logical connective} $\alpha$ is a pair
$\alpha = (\vz, \alpha_0)$ where:
\begin{enumerate}[$\bullet$]
\item $\vz = z_1, \dots, z_n$ is a sequence of distinct variables;
\item $\alpha_0 =\{a_1(\vec{x}_1), \dots, a_k(\vec{x}_k)\}$ is
a finite set of negative actions such that:
\begin{enumerate}[$-$]
\item the names $a_1,\ldots,a_k$ are distinct;
\item $\{\vec{x}_i\} \subseteq \{z_1,\ldots,z_n\}$ for each $1\leq i \leq k$.
    \end{enumerate}
\end{enumerate}
Two logical connectives are identified if one is obtained from another
by renaming of variables.
\end{defi}


We can   intuitively  explain the structure  of  logical connectives
in terms of standard connectives of
linear logic as follows.

The variables $z_1,\ldots,z_n$ play the role of
{\em placeholders} for (immediate) subformulas,
while $\alpha_0$ determines the \emph{logical structure} of $\alpha$.
An action $a(x_1,\ldots,x_m) \in \alpha_0$ can be seen as a kind of $m$-ary ``tensor product" $x_1 \otimes \cdots \otimes x_m$
indexed by the  name $a$.
  The whole set $\alpha_0$ can be  thought of as $k$-ary ``additive sum"
of its  elements:
$$\overbrace{\cdots \oplus \underbrace{(x_1 \otimes \cdots \otimes x_m)}_{a} \oplus \cdots}^{k \mbox{ components }}.$$

  In Appendix \ref{pol LL} we give a more precise
correspondence between   logical connective in our sense
and connectives of
polarized linear logic \cite{DBLP:journals/apal/Laurent04}.

\begin{exa} \label{example}
Consider the logical connective $\alpha = (x,y,z,t,
\{a(x,y,t), b(t,x),c(y,x)\})$.
By the previous discussion, we can intuitively think of it as
$$  \underbrace{(x \otimes y \otimes t)}_a \ \oplus \ \underbrace{(t \otimes x)}_b \ \oplus \ \underbrace{(y \otimes x)}_c.$$
When $\alpha$ is applied to  $\sN,\sM,\mathsf{K},\mathsf{L}$, it gives the formula
$$  \underbrace{(\sN \otimes \sM \otimes \mathsf{L})}_a \ \oplus \ \underbrace{(\mathsf{L} \otimes \sN)}_b
\ \oplus \ \underbrace{(\sM \otimes \sN)}_c.$$
\end{exa}



We now define behaviours built by logical connectives.

\begin{defi}[Behaviours defined by  logical connectives] \label{d-log conn beh}
Given an $m$-ary name $a$, an $n$-ary logical connective
$\alpha=(\vz, \alpha_0)$ with $\vz= z_1, \dots, z_n$ and behaviours
$\bN_1,\ldots,\bN_n,\bP_1,\ldots,\bP_n$ we define:
\begin{enumerate}[$\bullet$]
\item
$\coa \langle\bN_1, \ldots, \bN_m\rangle  :=
\{x_0|\coa \langle N_1, \ldots, N_m \rangle:
N_1 \in \bN_1,
\dots, N_m \in \bN_m\},$
\item
$\coal \langle \bN_1, \ldots, \bN_n \rangle  :=
\left(\bigcup_{a(\vx)\in\alpha_0} \coa\langle\bN_{i_1}, \dots, \bN_{i_m}\rangle\right)^{\bot\bot},$

\item

$\alpha(\bP_1,\ldots, \bP_n)  :=
\coal \langle \bP_1^\bot,\ldots, \bP_n^\bot \rangle\b,$
\end{enumerate}
where the indices $i_1, \dots, i_m \in \{1, \dots, n\}$ vary for each
$a(\vx)\in \alpha_0$ and
are determined by the variables $\vx = z_{i_1}, \dots, z_{i_m}$.
We  call the set
$$\coal_{\mathsf{eth}} \langle \bN_1, \ldots, \bN_n \rangle:= \bigcup_{a(\vx)\in\alpha_0} \coa\langle\bN_{i_1}, \dots, \bN_{i_m}\rangle$$
the \emph{ethics} of $\coal \langle \bN_1, \ldots, \bN_n \rangle$.
\end{defi}


\begin{rem} \label{ethics}
An ethics is  a set  of atomic  predesigns which are by construction linear in $x_0$.
It can be seen as a  ``generator" of  a behaviour defined
 by logical connectives in the following sense. For positives, we have
by definition $\coal  \langle \bN_1, \ldots, \bN_n \rangle = \coal_{\mathsf{eth}} \langle \bN_1, \ldots, \bN_n \rangle \b\b$.
For negatives, we have
by Proposition \ref{closure operator} (3):
$$\begin{array}{rcl}
\alpha(\bP_1,\ldots, \bP_n) & = &
 \coal \langle \bP_1^\bot,\ldots, \bP_n^\bot \rangle\b \\
 & = &  \coal_{\mathsf{eth}} \langle \bP_1^\bot,\ldots, \bP_n^\bot \rangle\b\b\b \\
 & = &  \coal_{\mathsf{eth}} \langle \bP_1^\bot,\ldots, \bP_n^\bot \rangle\b. \\
 &&\\
\end{array}$$
\end{rem}

\begin{exa}
Let $\alpha$ be the logical connective  as given  in Example
\ref{example} and $\bN,\bM,\bK,\mathbf{L}$
negative behaviours.
We have
$
\coal_{\mathsf{eth}}\langle\bN,\bM,\bK,\mathbf{L}\rangle  =  \coa \langle \bN,\bM,\mathbf{L} \rangle \;\cup\; \cob \langle \mathbf{L},\bN \rangle  \;\cup\; \coc \langle \bM,\bN \rangle
$.

\end{exa}

\begin{exa}[Linear logic connectives] \label{linear}
Logical connectives $\llpar, \with, \uparrowd,\bot, \top$
can be defined if the signature $\mathcal{A}$ contains
a nullary name $*$, unary names $\uparrow, \pi_1, \pi_2$ and
a binary name $\wp$. We also give notations to their duals
for readability.
$$
\begin{array}{rclrclrcl}
\pmb{\parr} & := & (x_1,x_2, \{\wp(x_1,x_2)\} ), &
\pmb{\otimes} & := & \overline{\pmb{\parr}},\quad &
\bullet & := & \overline{\wp},\\
\pmb{\with} & := & (x_1, x_2, \{\pi_1(x_1),\pi_2(x_2)\}),\qquad &
\pmb{\oplus} & := & \overline{\pmb{\with}}, &
\iota_i & := & \overline{\pi_i}, \\
\buparrowd & := & (x, \{\uparrowd(x)\}), &
 \bdownarrowd & := & \overline{\buparrowd},
& \downarrowd & := & \overline{\uparrowd}, \\
 \botto & := & (\epsilon,\{*\}), & \llone & := & \overline{\botto},\\
\toppo & := & (\epsilon, \emptyset), & \llzero & := & \overline{\toppo},
\end{array}
$$
where $\epsilon$ denotes the empty sequence.
We do not have exponentials here, because we are working
in a  nonlinear setting so that
they are already incorporated into the connectives.
With these logical connectives
we can build behaviours corresponding to usual linear logic types
(we use infix notations such as $\bN \pmb{\otimes} \bM$ rather than
the prefix ones $\pmb{\otimes} \langle \bN,\bM \rangle$).
$$
\begin{array}{rclrcl}
\bN \pmb{\otimes} \bM & = & \bullet \langle \bN,\bM \rangle\b\b,     & \bP \pmb{\parr} \bQ
 & = & \bullet \langle \bP\b,\bQ\b \rangle\b,  \\
\bN \pmb{\oplus} \bM & = & (\iota_1 \langle \bN \rangle \cup \iota_2 \langle \bM \rangle)\b\b,\qquad     &
\bP \pmb{\with} \bQ  & = & \iota_1 \langle \bP\b \rangle \b \cap \iota_2 \langle \bQ\b \rangle\b, \\
\bdownarrowd \bN & = & \downarrowd \langle \bN \rangle\b\b,     &
\buparrowd \bP & = & \downarrowd \langle\bP\b \rangle\b,   \\
\mathbf{1} & = & \{x_0|\overline{*}\}\b\b, &
\botto & = & \{x_0|\overline{*}\}\b,\\
\mathbf{0} & = & \emptyset\b\b, &
\toppo & = & \emptyset\b. \\
&&&&&\\
\end{array}
$$
\end{exa}

The next theorem
illustrates a special feature of behaviours
defined by logical connectives.
It also suggests that
nonlinearity and universal nondeterminism play dual roles.

\begin{thm} \label{p-duphalf} \label{fundI} \label{p-log}
Let $\bP$ be an arbitrary positive behaviour.
\begin{enumerate}[\em(1)]
\item $P \models x_1: \bP, \ x_2: \bP\ \Longrightarrow \
P[x_0/x_1,x_0/x_2] \in  \bP$.
\item $ \Bigwedge \bX \in \bP\b \ \Longrightarrow\  \bX \subseteq \bP\b$.
\end{enumerate}
Moreover, if  $\bP$ is \emph{obtained by applying a logical connective}, that is $\bP = \coal\langle \bN_1,\ldots,\bN_n\rangle$ for some $\alpha$, $\bN_1,\ldots,\bN_n$,  then:
\begin{enumerate}[\em(1)]
\setcounter{enumi}{2}
\item the converse of {\em{(1)}} (\emph{duplicability}) and
 \item the converse of {\em{(2)}} (\emph{closure under $\Bigwedge$})  hold.
\end{enumerate}
\end{thm}

\proof\hfill
\begin{enumerate}[(1)]
\item For any $N\in \bP^\bot$, we have $\pl P[N/x_1, N/x_2]\pr=\maltese$.
Hence, $\pl P[x_0/x_1,x_0/x_2][N/x_0] \pr = \maltese$, and so
$P[x_0/x_1,x_0/x_2] \in \bP^{\bot\bot} = \bP$.
\item By Proposition \ref{conj and ort} (3), we have $\Bigwedge \bX \preceq \Bigwedge \{N\} = N$
for any $N \in \bX$. Since $\bP\b$ is a behaviour,
it is upward closed with respect to $\preceq$.
Hence the claim holds.
\item[(4)]
For the sake of  readability, we consider
the binary case and  show
that $N, M \models \bP\b$ implies $N \wedge M \models \bP\b$.
 The
general case can be proven using the same argument.

Let $ \bP\b = \coal\langle \bN_1,\ldots,\bN_n \rangle^\bot
= \alpha (\bN_1^\bot,\ldots,\bN_n^\bot)$.
To prove $N \wedge M \in \bP\b$, by Remark \ref{ethics}, it is sufficient to show that
$N \wedge M $ is orthogonal to any
$x_0 |\coa\langle \vec{K}\rangle \in \coal_{\mathsf{eth}}\langle \bN_1,\ldots,\bN_n \rangle$.
Since by construction  $x_0$ occurs only once at the head position of $x_0 |\coa\langle \vec{K}\rangle$,
we only have to show that
$\pl N \wedge M  \;|\; \coa\langle \vec{K}\rangle \pr = \maltese$.


Let $N=\Sum a(\vx).P_a$ and  $M= \Sum a(\vx).Q_a$
so that $N \wedge M = \Sum a(\vx).(P_a \wedge Q_a)$.
Since $N \wedge M  \;|\; \coa \langle \vec{K} \rangle$
is a predesign, we have by Lemma \ref{l-facts} (2), (3):
$$ \pl N \wedge M  \;|\; \coa \langle \vec{K} \rangle \pr =
  \pl P_a \wedge  Q_a[\vec{K}/\vx]\pr =
  \pl P_a[\vec{K}/\vx]\pr \wedge  \pl Q_a[\vec{K}/\vx]\pr
= \pl N \;|\; \coa \langle \vec{K} \rangle \pr \wedge
\pl M \;|\; \coa \langle \vec{K} \rangle \pr.$$
Since $N,M \in \bP\b$, we have
$\pl N \;|\; \coa \langle \vec{K} \rangle \pr = \maltese$
and $ \pl M \;|\; \coa \langle \vec{K} \rangle \pr =\maltese$.
Our claim then immediately follows.
\item[(3)] Let
$P[x_0/x_1, x_0/x_2]\in \bP
=\coal\langle \bN_1,\ldots,\bN_n \rangle$. It suffices to show that
$\pl P[N/x_1, M/x_2]\pr=\maltese$ holds for any $N, M \in \bP\b$.
But we have just proven that $N\wedge M \in \bP\b$, and so
$\pl P[x_0/x_1, x_0/x_2][N\wedge M/x_0] \pr = \pl P[N\wedge M/x_1,
N\wedge M/x_2] \pr = \maltese$.
Since $N\wedge M \preceq N, M$ by Proposition \ref{conj and ort} (3),
we have $\pl P[N /x_1,M/x_2] \pr = \maltese$.
\qed
\end{enumerate}

\begin{rem}
Theorem \ref{p-log} can be considered as an internal,
monistic form of  soundness and completeness for the
contraction rule: soundness
corresponds to point (1) while completeness
to its converse (3), duplicability.

However, in the sequel we only use point (1) (in Theorem \ref{soundness}) and point (4) (in Lemma \ref{l-comp})
of Theorem \ref{p-log}.
\end{rem}


\subsection{Internal completeness}\label{ss-internal}

In \cite{DBLP:journals/mscs/Girard01}, Girard proposes
a purely monistic, local notion of completeness,
called {\em internal completeness}.
It  means that   we can give a precise
and direct description to the elements of behaviours (built by
logical connectives) without using
the orthogonality and without referring to any proof system.
It is easy to see that
negative logical connectives enjoy internal completeness:

\begin{thm}[Internal completeness (negative case)]\label{t-negative}
Let $\alpha = (\vz,\alpha_0)$ be a logical connective with
$\vz = z_1,\ldots,z_n$ and $N =\Sum a(\vx).P_a$ an atomic negative design.  We have:
$$
\begin{array}{ccl}
N \in \alpha(\bP_1,\ldots, \bP_n)  & \Longleftrightarrow
&  P_a  \models z_{i_1} : \bP_{i_1}, \ldots, z_{i_m} : \bP_{i_m},
\mbox{ for every $a(\vx)\in\alpha_0$},
\end{array}
$$
where the indices $i_1, \dots, i_m \in \{1,\ldots,n\}$ are determined by the variables
$\vx= z_{i_1}, \dots, z_{i_m}$.
\end{thm}
\proof
Let  $N=\Sum a(\vx).P_a$ be an  atomic  negative design and
$P = x_0| \coa \langle N_1,\ldots,N_m \rangle \in
\coal_{\mathsf{eth}} \langle\bP_1^\bot,\ldots, \bP_n^\bot \rangle
=\bigcup_{a(\vx)\in\alpha_0} \coa\langle\bP_{i_1}^\bot, \dots, \bP_{i_m}^\bot\rangle$.
Since $P[N/x_0]$ is a predesign and
$x_0$ occurs only at the head position of $P$, we have
by Lemma \ref{l-facts} (2):
$$ \pl P[N/x_0] \pr = \pl \Sum a(\vx).P_a  \;|\; \coa \langle N_1,\ldots,N_m \rangle \pr
= \pl
P_a[N_1/z_{i_1},\ldots,N_m/z_{i_m}] \pr.$$
This means that $N \in
\coal_{\mathsf{eth}} \langle\bP_1^\bot,\ldots, \bP_n^\bot \rangle \b
= \alpha(\bP_1,\ldots,\bP_n)$ (see Remark \ref{ethics})
  if and only if for  every $a(\vx) \in \alpha_0$
and for every $ N_1 \in \bP_{i_1}^\bot,\ldots, N_m \in \bP_{i_m}^\bot$,
$\pl P_a[N_1/z_{i_1},\ldots,N_m/z_{i_m}]\pr=\maltese $
 if and only if
  for  every $a(\vx) \in \alpha_0$,
 $P_a  \models z_{i_1} : \bP_{i_1}, \ldots, z_{i_m} : \bP_{i_m}$
(see Definition \ref{d-context} (b)).
\qed
Notice that in the above, $P_b$ can be arbitrary
 when
$b(\vec{y})\notin\alpha_0$. Thus our approach is ``immaterial''
in that we do not consider  material designs (see \eg
\cite{DBLP:journals/mscs/Girard01,DBLP:journals/corr/abs-cs-0501039,Terui08} for the  definition of material design).
The original ``material'' version of internal completeness
\cite{DBLP:journals/mscs/Girard01}
can be easily derived from our immaterial one.

\begin{rem}
A remarkable example of internal completeness for negative behaviours is provided
for  the logical connective $\pmb{\with}= (x_1, x_2, \{\pi_1(x_1),\pi_2(x_2)\})$:
$$
\begin{array}{ccl}
N \in \bP \pmb{\with} \bQ \!\! & \Longleftrightarrow
& N= \pi_1(x_1). P + \pi_2(x_2). Q + \cdots \mbox{ , for some  }  P\models x_1 : \bP
\mbox{ and }
Q \models x_2 : \bQ \\
& \Longleftrightarrow & N = \pi_1(x_0). P + \pi_2(x_0). Q + \cdots \mbox{ , for some  } P\in \bP
\mbox{ and }
 Q \in \bQ.
\end{array}
$$
Above, the irrelevant components of the sum are suppressed by ``$\cdots$.''
Up to materiality  (\ie  removal of irrelevant additive components),
$\bP \pmb{\with} \bQ$, which has been defined by \emph{intersection},
is isomorphic to
the \emph{cartesian product} of $\bP$ and $\bQ$. This
 isomorphism is  called
``\emph{the mystery of incarnation}" in \cite{DBLP:journals/mscs/Girard01}.
\end{rem}

As to positive connectives,
\cite{DBLP:journals/mscs/Girard01} proves internal completeness
theorems
for additive and multiplicative ones separately in the linear and
deterministic setting. They are
integrated in \cite{Terui08} as follows:

\begin{thm}[Internal completeness (linear, positive case)]
When the universe of standard designs is restricted to linear and
deterministic ones,
we have
$$\coal \langle \bN_1, \ldots, \bN_n \rangle =
\coal_{\mathsf{eth}} \langle \bN_1, \ldots, \bN_n \rangle \cup\{\maltese\}.$$
\qed
\end{thm}

However, this is no more true with nonlinear designs.
A counterexample is given below.

\begin{exa} \label{example completeness}
Let us consider the behaviour
$\bP:= \bdownarrowd \langle\buparrowd(\zero) \rangle
= \bdownarrowd_{\mathsf{eth}} \langle\buparrowd(\zero) \rangle\b\b$
and the designs $P =  x_0|\mathord{\downarrow} \langle \mathord{\uparrowd}(y).\maltese \rangle$
and $Q = x_0|\mathord{\downarrow} \langle \mathord{\uparrowd}(y).P \rangle$
of Remark \ref{r-separation}.
By construction,
$P$
belongs
to $\bP$. Since $ P \preceq Q$,
$Q$ also belongs to $\bP$.
However, $Q\not\in
\bdownarrowd_{\mathsf{eth}} \langle\buparrowd(\zero) \rangle$,
since $\uparrowd(y).P$ is not atomic and so
cannot belong to $\buparrowd(\zero)$.
\end{exa}

This motivates us to directly prove completeness for proofs,
rather than deriving it from internal completeness as in the
original work \cite{DBLP:journals/mscs/Girard01}.

In \cite{Basaldella09} a weaker form of internal completeness is
proved, which is enough to derive a weaker form of full completeness:
all {\em finite} ``winning'' designs are interpretations
of proofs. While
such a finiteness assumption is quite common in game semantics,
we will show that it can be avoided in ludics.

We end this section with the following remark.
\begin{rem}
The main linear logic isomorphism, namely the exponential one
$ \oc A \otimes \oc B \ \cong \ \oc(A \with B)$ can be expressed
in our notation as
$ \buparrowd \bP \pmb{\otimes} \buparrowd \bQ \ \cong \ \bdownarrowd ( \bP \pmb{\with} \bQ )$.

 In our setting it is possible to prove that  those behaviours
are ``morally" isomorphic, in the sense that
they are isomorphic if we consider designs equal
up to materiality\footnote{Informally,
two designs $D$ and $E$ are equal up to materiality in a behaviour
$\bG$ if they only differ
in  occurrences of positive subdesign
which are  irrelevant for the normalization against designs
of $\bG\b$.}.

We can in fact  define a pair of maps $(f,g)$ on designs
 such that:
\begin{enumerate}[$\bullet$]
\item
$f: \ \buparrowd \bP \pmb{\otimes} \buparrowd \bQ \ \longrightarrow \ \bdownarrowd ( \bP \pmb{\with} \bQ )$ and  $
g : \ \bdownarrowd ( \bP \pmb{\with} \bQ ) \ \longrightarrow \ \buparrowd \bP \pmb{\otimes} \buparrowd \bQ$;

\item if $P$ and $Q$ are equal
up to materiality in $\buparrowd \bP \pmb{\otimes} \buparrowd \bQ$, then $f(P)$ and $f(Q)$ are equal up to materiality in $\bdownarrowd ( \bP \pmb{\with} \bQ )$,
and similarly for $g$;
 \item
for any $P \in \buparrowd \bP \pmb{\otimes} \buparrowd \bQ$, we have that
 $g (f  (P))$ and $P$ are equal  up to materiality in
$\buparrowd \bP \pmb{\otimes} \buparrowd \bQ$,
and similarly for the other direction.
\end{enumerate}
\end{rem}

\medskip\noindent We postpone a detailed study of    isomorphisms of types  and related issues  to a subsequent work.

\section{Proof system and completeness for proofs}\label{s-full}

\noindent Having set up the framework, we now address the main problem: an
interactive form of G\"{o}del completeness.
We first introduce the proof system in \ref{proof system},
then examine its soundness in \ref{ss-soundness}, and finally prove
completeness in \ref{completeness}, in a way quite analogous to the
proof of G\"odel's theorem based on proof search
(often attributed to Sch\"utte \cite{Schutte56}).

\subsection{Proof system} \label{proof system}
We will now introduce a proof system. In our system,
logical rules are automatically generated by
logical connectives. Since  the names which constitute the
logical connectives
are chosen among the names of a signature $\cA$,
the set of logical connectives vary
for each signature $\mathcal{A}$. Thus,
 our proof system is parameterized
by $\mathcal{A}$.

 If one chooses  $\mathcal{A}$ rich enough,
the constant-only fragment of polarized linear logic
(\cite{DBLP:journals/apal/Laurent04}; see also
\cite{DBLP:journals/corr/abs-cs-0501039})
can be embedded, as we will show in Appendix
\ref{pol LL}.

In the sequel, we focus on \emph{logical} behaviours, which
are composed by using logical connectives only.
\begin{defi}[Logical behaviours] \label{logical behaviours}
A behaviour is \textbf{logical} if it is inductively
built as follows ($\alpha$ denotes an arbitrary logical connective):
$$
\bP  :=  \coal\langle \bN_1, \dots, \bN_n\rangle, \qquad
\bN  :=  \alpha(\bP_1, \dots, \bP_n).
$$
\end{defi}
\noindent Notice that the orthogonal of a logical behaviour is again logical.

As advocated in the introduction,
our monistic framework renders both proofs and models as
homogeneous objects: designs.

\begin{defi}[Proofs, Models] \label{proof,models}
A \textbf{proof} is a standard design (Definition \ref{d-standard})
in which all the conjunctions  are  unary.
In other words, a proof is  a total, deterministic and $\maltese$-free design
without cuts and identities.
A \textbf{model} is a linear standard design
(in which conjunctions of arbitrary cardinality may occur).
\end{defi}

We will use proofs as proof-terms for syntactic derivations
in the proof system to be introduced below. In that perspective,
it is reasonable to exclude designs with non-unary conjunctions from proofs,
because
they do not have  natural counterparts in logical reasoning. For instance,
the nullary conjunction (daimon) and the binary one would correspond to
the following ``inference rules'' respectively:
$$
\infer{\vdash \Gamma}{} \qquad\qquad
\infer{\vdash \Gamma}{\vdash \Gamma & \vdash \Gamma}
$$
with $\vdash \Gamma$ an arbitrary sequent. Notice that
we have not specified yet what a proof actually proves.
Hence
it might be better called  ``proof attempt'' or ``untyped proof''
or ``para-proof.''


On the other hand, we restrict models to linear designs
just to emphasize the remarkable fact
that \emph{linear} designs do suffice
for defeating any failed proof attempt that is possibly \emph{nonlinear}.

Given a  design $D$, let $\mathsf{ac}^+(D)$ be the set of occurrences
of  positive actions $\coa$ in $D$. The {\em cardinality} of $D$
is defined to be the cardinality of $\mathsf{ac}^+(D)$.
For instance, the fax
$\eta(x) = \Sum a(y_1,\ldots,y_n).x|\coa \langle \eta(y_1),\ldots,\eta(y_n) \rangle$ (see Section \ref{ortho-s})
is an infinite  design in this sense.
Also, both proofs and models can be infinite. 

A {\em positive} (resp.\ {\em negative}) {\em sequent} is a pair of the form
$P\vdash \bGamma$ (resp.\ $N\vdash\bGamma, \bN$)
where $P$ is a positive proof
(resp.\ $N$ is a negative proof) and
 $\bGamma$ is a positive context of logical behaviours (Definition \ref{d-context} (a))
such that  $\mathsf{fv}(P)\subseteq \mathsf{fv}(\bGamma)$
(resp.\ $\mathsf{fv}(N)\subseteq \mathsf{fv}(\bGamma)$).

We write $D \vdash \mathbf{\Lambda}$ for a generic sequent.
Intuitively, a sequent $D \vdash \mathbf{\Lambda}$  should be understood as a claim that
``$D$ is a proof of $ \vdash \mathbf{\Lambda}$'' or ``$D$ is of type $ \vdash \mathbf{\Lambda}$.''

Our proof system consists of two sorts of inference rules:
\begin{enumerate}[$\bullet$]
\item A {\em positive} rule $(\coal, \coa)$:

$$
\infer[(\coal, \coa)]{z|\coa\langle M_1, \dots, M_m
\rangle\vdash
\bGamma}{
M_1 \vdash \bGamma, \bN_{i_1} &
\dots &
M_m \vdash \bGamma, \bN_{i_m}
& (z:\coal\langle\bN_1, \dots, \bN_n\rangle \in \bGamma)}$$
where $\alpha = (\vz, \alpha_0)$, $\vz = z_1, \dots, z_n$ and
$a(\vec{x})\in \alpha_0$ so that the indices
$i_1, \dots, i_m \in \{1, \dots,  n\}$ are determined by the variables
$\vec{x} = z_{i_1}, \dots, z_{i_m}$.

\item
A {\em negative} rule $(\alpha)$:
$$
\infer[(\alpha)]{\Sum a(\vx).P_a \vdash
\bGamma, \alpha(\bP_1, \dots, \bP_n)}{
\{P_a \vdash \bGamma, z_{i_1}: \bP_{i_1}, \dots, z_{i_m}: \bP_{i_m}\}_{a(\vx)\in \alpha_0}}
$$
where, as in the positive rule,
the indices
$i_1, \dots, i_m$ are determined by the variables
$\vec{x} = z_{i_1}, \dots, z_{i_m}$ for each $a(\vx)\in \alpha_0$.

We assume that $\vx$ are fresh, \ie do not occur in
$\bGamma$. This does not cause a loss of generality
since variables in $\alpha$ can be renamed (see Definition \ref{d-lc}).

Notice that a component $b(\vy).P_b$ of
$\Sum a(\vx).P_a$ can be arbitrary when
$b(\vec{y})\not\in \alpha_0$.
Hence we again take
an ``immaterial'' approach (\cf Theorem \ref{t-negative}).

\end{enumerate}

\medskip\noindent Observe that the positive rule
$(\coal, \coa)$ involves implicit uses of the contraction rule
on positive behaviours.
The weakening rule for positive behaviours is implicit too; in
the bottom up reading of a proof derivation, unused formulas
are always propagated to the premises of any instance of rule.
It should also be noted that proof search in our system is
deterministic. In particular, given a positive sequent
$z|\coa\langle M_1, \dots, M_m \rangle\vdash\bGamma$,
the head variable $z$ and the first positive action $\coa$
completely determine the next positive rule
to be applied bottom-up (if there is any).

It is also possible to adopt a ``material'' approach in the proof system
by simply requiring $P_b = \Omega$
when $b(\vec{y})\not\in \alpha_0$ in the rule $(\alpha)$.
Then a proof $D$ is finite
(\ie $\mathsf{ac}^+(D)$ is a finite set)
whenever $D\vdash \mathbf{\Lambda}$ is derivable for some $ \mathbf{\Lambda}$. Thus,
as in ordinary sequent calculi, our proof system accepts only
{\em essentially finite} proofs for derivable sequents
(\ie finite up to removal of irrelevant parts).

\begin{rem} \label{example top fax}
To clarify the last point, we  observe
that for any (\emph{possibly infinite}) negative proof
$N$ with $\mathsf{fv}(N) \subseteq \mathsf{fv}(\bGamma)$, the sequent  $N \vdash \bGamma,\toppo$ is derivable
by the instance of the negative rule with
$\alpha= \toppo = (\epsilon,\emptyset)$. In fact, this
corresponds to the usual top-rule of linear logic
 (see also Example \ref{linear rules}):
\begin{prooftree}
\AxiomC{} \RightLabel{$(\toppo)$}
\UnaryInfC{$N \vdash \bGamma,\toppo$}
\end{prooftree}
This means that  for a (possibly infinite) negative proof $N$
there is a finite derivation of $N \vdash \bGamma,\toppo$.
By contrast, in the ``material'' approach
we only have
\begin{prooftree}
\AxiomC{} \RightLabel{$(\toppo)$}
\UnaryInfC{$\Sum a(\vx).\Omega \vdash \bGamma,\toppo$}
\end{prooftree}
where $\Sum a(\vx).\Omega $ is the unique  negative proof which has cardinality $0$.
\end{rem}

\begin{exa} \label{linear rules}
For linear logic connectives (Example \ref{linear}), the positive and negative rules
specialize to the following (taking here the ``material'' approach):

$$
\begin{array}{c}
\infer[ (\pmb{\lltensor}, \bullet)]{z|\bullet\langle M_1, M_2\rangle \vdash
\bGamma}{
M_1 \vdash \bGamma, \bN_1 &
M_2 \vdash \bGamma, \bN_2 &
(z:\bN_1 \pmb{\otimes} \bN_2 \in \bGamma)}
\quad
\infer[\pmb{\llpar}]{\wp(x_1, x_2). P  \vdash \bGamma,
\bP_1 \pmb{\llpar} \bP_2}{P \vdash \bGamma, x_1 : \bP_1, x_2 : \bP_2}\\[1em]
\infer[(\pmb{\llplus}, \iota_i)]{z|\iota_i\langle M\rangle \vdash
\bGamma}{
M \vdash \bGamma, \bN_i & (z:\bN_1 \pmb{\llplus} \bN_2 \in \bGamma)}
\qquad
\infer[\pmb{\llwith}]{\pi_1(x_1). P_1 + \pi_2(x_2). P_2  \vdash \bGamma,
\bP_1 \pmb{\llwith} \bP_2}{P_1 \vdash \bGamma, x_1 : \bP_1 &
P_2 \vdash \bGamma, x_2 : \bP_2}\\[1em]
\infer[(\bdownarrowd, \downarrowd)]{z|\downarrowd\langle N\rangle \vdash
\bGamma }{
N \vdash \bGamma, \bN & (z: \bdownarrowd \bN \in \bGamma) }
\qquad
\infer[\buparrowd]{\uparrowd(x).P  \vdash \bGamma,
\buparrowd\bP}{P \vdash \bGamma, x :\bP}\\[1em]
\infer[(\llone,\overline{*})]{z|\overline{*}\vdash\bGamma}{
 (z:\llone \in \bGamma)}
\qquad
\infer[(\botto)]{*.P\vdash\bGamma,\botto}{P\vdash\bGamma}
\qquad
\infer[(\toppo)]{\Sum a(\vx).\Omega \vdash\bGamma,\toppo}{}
\end{array}
$$
\end{exa}
\subsection{Soundness}\label{ss-soundness}

The inference rules given above are all sound. Namely we have:

\begin{thm}[Soundness] \label{soundness}
If $D \vdash \mathbf{\Lambda}$ is derivable in the proof system, then
$D  \models \mathbf{\Lambda}$.
\end{thm}
\proof
By induction on the length of the derivation of
$D \vdash \mathbf{\Lambda}$. We have two cases,
one for each sort of rule.
\begin{enumerate}[(1)]
\item Suppose that
the last inference rule is

$$
\infer[(\coal, \coa)]{z|\coa\langle M_1, \dots, M_m
\rangle\vdash
\bGamma}{
M_1 \vdash \bGamma, \bN_{i_1} &
\dots &
M_m \vdash \bGamma, \bN_{i_m}
& (z:\coal\langle\bN_1, \dots, \bN_n\rangle \in \bGamma)}$$
where $\bGamma = x_1 : \bP_1, \dots, x_l: \bP_l$
and
$z:\coal\langle\bN_1, \dots, \bN_n\rangle = x_k : \bP_k$
for some $1\leq k\leq l$.

The induction hypothesis gives us
$M_j \models \bGamma, \bN_{i_j}$ for every $1\leq j\leq m$.
By Lemma \ref{l-closure} (3),
$M_j' := \pl  M_j [N_1/x_1,\ldots,N_l/x_l]\pr \in \bN_{i_j}$ for
every $N_1 \in \bP_1^\bot, \dots, N_l\in \bP_l^\bot$ and
by Definition \ref{d-lc}, we have that
$x_0|\coa\langle M_1', \dots, M_m'\rangle \in
\coal\langle\bN_1, \dots, \bN_n\rangle$, that is
$x_0|\coa\langle M_1', \dots, M_m'\rangle \models
x_0 : \coal\langle\bN_1, \dots, \bN_n\rangle$.

Applying  Lemma \ref{l-closure} (1), we get
$ x_0|\coa\langle   M_1   , \dots,   M_m   \rangle  \models \bGamma, x_0 :
\coal\langle\bN_1, \dots, \bN_n\rangle$
and
by Theorem \ref{p-duphalf} (1)
we conclude $ z|\coa\langle   M_1   , \dots,   M_m  \rangle  \models \bGamma$.
\item Suppose now that
the last inference rule is

$$
\infer[(\alpha)]{\Sum a(\vx).P_a \vdash
\bGamma, \alpha(\bP_1, \dots, \bP_n)}{
\{P_a \vdash \bGamma, \vx: \vec{\bP}_a\}_{a(\vx)\in \alpha_0}}
$$
where $\bGamma = y_1 : \bQ_1, \dots, y_l: \bQ_l$
and $\vx : \vec{\bP}_a$ stands
for $ z_{i_1}: \bP_{i_1}, \dots, z_{i_m}: \bP_{i_m}$.
We assume that the  variables $y_1,\ldots,y_l$ and $\vx$
are disjoint in any premise.

The induction hypothesis gives us
$P_a \models \bGamma, \vx: \vec{\bP}_a$ for every $a(x) \in \alpha_0$.
By Lemma \ref{l-closure} (1),  for
every $N_1 \in \bQ_1^\bot, \dots, N_l\in \bQ_l^\bot$,
$P_a' := \pl  P_a [N_1/x_1,\ldots,N_l/x_l]\pr \models \vx: \vec{\bP}_a$.

Then, we can apply  Theorem \ref{t-negative} to obtain
$\Sum a(\vx).P_a' \in \alpha(\bP_1, \dots, \bP_n)$, that is
$\Sum a(\vx).P_a' \models \alpha(\bP_1, \dots, \bP_n)$.
Notice that in $\Sum a(\vx).P_a'$
the components $b(\vy).P_b$ for $b(\vy) \notin \alpha_0$
can be arbitrary.

We finally apply Lemma \ref{l-closure} (3) and conclude
$ \Sum a(\vx).P_a \models
\bGamma, \alpha(\bP_1, \dots, \bP_n)$.
\qed
\end{enumerate}

\medskip\noindent Although our proof system does not include a cut
rule officially, the semantics validates it as follows.

\begin{prop} \hfill
\begin{enumerate}[\em(1)]
\item If $P \models \bGamma, z: \bP$ and
$M \models \bGamma,\bP\b$,
then $\pl P[M/z] \pr \models \bGamma$.
\item If $N \models \bN,\bGamma, z: \bP$ and
$M \models \bGamma,\bP\b$,
then $\pl N[M/z] \pr \models \bN,\bGamma$.
\end{enumerate}
\end{prop}
\proof
Let $\bGamma$ be $x_1 : \bP_1,\ldots, x_n : \bP_n$,
let $K_1 \in \bP_1^\bot,\ldots,K_n \in \bP_n^\bot$ and write $\vec{K}/\vx$ for $K_1/x_1,\ldots,K_n/x_n$.
\begin{enumerate}[(1)]
\item
By Lemma \ref{l-closure}, we have $P':= \pl  P[\vec{K}/\vx] \pr \models z: \bP$
and $M':=  \pl M[\vec{K}/\vx] \pr \models \bP\b$,
so that  $P'[x_0/z] \in \bP$ and
and $M' \in \bP\b$.
Hence,
$ \pl P'[x_0/z][M'/x_0] \pr = \pl P'[M'/z] \pr= \maltese$.
From this fact and
associativity (Theorem \ref{t-assoc}), we can derive
 $\pl   \pl P[M/z] \pr [\vec{ K }/\vx]   \pr =\maltese$,
which proves $\pl P[M/z] \pr \models \bGamma$.

\item Let $Q$ be an arbitrary design in $\bN\b$.
By Lemma \ref{l-closure}, we obtain  $\pl  Q[N/x_0] \pr \models \bGamma, z: \bP$
and  $Q' = \pl \pl  Q[N/x_0] \pr [\vec{K}/\vx] \pr
\models z: \bP$. On the other side, we have
$M':=  \pl M[\vec{K}/\vx] \pr \models \bP\b$.
From   $Q'[x_0/z] \in \bP$ and
and $M' \in \bP\b$,
we obtain
$ \pl Q'[x_0/z][M'/x_0] \pr = \pl Q'[M'/z] \pr= \maltese$.
From this fact and
associativity, we can derive
 $\pl   Q[ \; \pl N[M/z] \pr[\vec{ K }/\vx] \; /x_0]   \pr =\maltese$,
which proves $\pl N[M/z] \pr \models \bN,\bGamma$. \qed
\end{enumerate}

\medskip\noindent Thanks to the previous proposition, we can naturally
strengthen our proof system as follows.  First, we consider sequents
of the form $D \vdash \mathbf{\Lambda}$ where $D$ is a ``proof with
cuts" (\ie a proof in the sense of Definition \ref{proof,models}
except that the cut-freeness condition is not imposed).
Second, we add  the following
cut rule:
$$
\infer[(cut)]{D[N/z]\vdash \mathbf{\Xi},\bGamma}{
D \vdash \mathbf{\Xi}, \bGamma, z:\bP
&
N \vdash \bGamma, \bP^\bot}$$
where $\mathbf{\Xi}$ is either empty
or it consists of a negative logical behaviour $\bN$.

The soundness theorem can  be naturally generalized  as follows:
\begin{thm}[Soundness (with cut rule)]
If $D \vdash \mathbf{\Lambda}$ is derivable
in the proof system  with the cut rule above,
then  $\pl D \pr  \models \mathbf{\Lambda}$. \qed
\end{thm}

\subsection{Completeness for proofs} \label{completeness}

Let us finally
establish the other direction of Theorem \ref{soundness}, namely:

\begin{thm}[Completeness for proofs] \label{full completeness}
A sequent
$D \vdash \bLambda$ is derivable in the proof system
if and only if
$D  \models \bLambda$.

In particular, for any positive logical behaviour $\bP$ and a proof $P$,
$P\vdash x_0:\bP$ is derivable if and only if $P \in \bP$.
Similarly for the negative case. \qed
\end{thm}

%

\noindent Before proving the theorem, let us recall a
well-established method for
proving G\"odel completeness
based on proof search
(often attributed to Sch\"utte \cite{Schutte56}).
It proceeds as follows:
\begin{enumerate}[(1)]
\item Given an unprovable sequent $\vdash \Gamma$,
find an open branch in the cut-free proof search tree.
\item From the open branch, build a countermodel $M$
in which $\vdash \Gamma$ is false.
\end{enumerate}

\noindent The proof below follows the same line of argument.
We can naturally adapt (1) to our setting, since
the bottom-up cut-free
proof search in our proof system is deterministic in the sense that
at most one rule applies at each step. Moreover,
it never gets stuck at the negative sequent,
since a negative rule is always applicable bottom-up. Adapting
(2) is more delicate.

For simplicity, we assume that
the sequent $D\vdash \bLambda$ is positive; the argument below
can be easily adapted to the negative case.
So, suppose that a positive sequent
$P_0 \vdash \bTheta_0$ with
$\bTheta_0 = x_1: \bP_1, \dots, x_n:\bP_n$
does not have a derivation.
By K\"{o}nig's Lemma,
there exists a branch $\mathsf{ob}$ in the cut-free proof search tree,
$$\begin{array}{ccc}
 \deduce{\phantom{A}}{
\deduce{\phantom{A}}{
\deduce{\mathsf{ob}}{
\deduce{}{}}}} &
 \deduce{\phantom{A}}{
\deduce{\phantom{A}}{
\deduce{=}{
\deduce{}{}}}}
 &
 \infer{P_0\vdash \bTheta_0}{
\infer{N_0\vdash \bPsi_0}{
\infer{P_1\vdash \bTheta_1}{
\infer*{N_1\vdash \bPsi_1}{}}}},
\end{array}
$$
which
is {\em either} finite and has the topmost sequent
$P_{max} \vdash \bTheta_{max}$ with $max \in\mathbb{N}$
to which no rule applies anymore,
{\em or} infinite. In the latter case, we set $max = \infty$.

Our goal is to build  models
$\cM(x_1) \in \bP_1^\bot,
\dots
\cM(x_n) \in \bP_n^\bot$
such that
$$\Norm{P_0[\cM(x_1)/x_1, \dots, \cM(x_n)/x_n]}= \Omega.$$
More generally, we define negative designs
\begin{enumerate}[$\bullet$]
\item $\cM(i)$ for every $i\geq 0$ ($0\leq i\leq max$ if $max\in \mathbb{N}$);
\item $\cM(x)$ for every variable $x$ occurring in the branch.
\end{enumerate}
Below, $\alpha$ and $\beta$ stand for
logical connectives: $\alpha = (\vz, \alpha_0)$, $\beta = (\vec{u}, \beta_0)$.

To define $\cM(i)$ we distinguish three cases:

\noindent(i)
When $i= max$ and
$P_{max} = \Omega$, let $\cM(max) := \maltese^- ( = \Sum a(\vec{x}).\maltese)$.

\noindent(ii) When $i = max$ and
$P_{max} \neq \Omega$,
suppose that $P_{max} \vdash \bTheta_{max}$ is of the form
$z|\coc\langle \vec{M}\rangle\vdash \bGamma, z: \coal \langle \vec{\bN}\rangle$
but $c(\vec{w}) \notin \alpha_0$ so that the proof search gets stuck.
Then let $\cM(max) := \Sum_{a(\vx) \in \alpha_0} a(\vx).\maltese$.
Recall that the partial sum $\cM(max)$ has
$c(\vec{w}).\Omega$ as component by our convention.

\noindent(iii) For $i<max$,
suppose that the relevant part of the branch $\mathsf{ob}$ is of the form:
$$
 \deduce{\phantom{A}}{
\deduce{\phantom{A}}{
\deduce{\phantom{a}}{
\deduce{\phantom{a}}{
\deduce{\mathsf{ob}}{
\deduce{}{}}}}}}
\quad\raisebox{3em}{$=$}
\quad
\infer*{}{
\infer{P_i \vdash \bTheta_i}{
    \infer{N_i \vdash \bPsi_i}{
        \infer*{P_{i+1}\vdash \bTheta_{i+1}}{}}}}
\quad\raisebox{3em}{$=$}
\quad
\infer*{}{
\infer[(\coal, \coa)]{z|\coa\langle M_1,\ldots,M_{{k-1}}, \Sum b(\vy).P_{i+1}, M_{k+1},\ldots,M_{{m}}\rangle\vdash \bTheta_i}{
    \infer[(\beta)]{\Sum b(\vy). P_{i+1} \vdash \bTheta_i, \bN_{i_k}}{
        \infer*{P_{i+1} \vdash \bTheta_i, y_1 : \bQ_1, \dots, y_l : \bQ_l}{}}}},
$$
where
$\bTheta_i$ contains $z: \coal \langle \bN_1,\ldots,\bN_n \rangle$,
$a(\vec{x}) \in \alpha_0$ with $\vec{x}= z_{i_1},\ldots,z_{i_m}$, and
$\bN_{i_k} = \beta(\bQ_1,\ldots,$ $\bQ_s)$,
 $b(\vy) \in \beta_0$ with $\vy=y_1,\ldots,y_l$.
Namely, the situation is as follows (to be read bottom-up):
\begin{enumerate}[$\bullet$]
\item The head variable of $P_i$ is $z$, so
$z: \coal \langle \bN_1,\ldots,\bN_n \rangle$ is chosen from
the context $\bTheta_i$
and the rule $(\coal, \coa)$ is applied. Among $m$ upper sequents,
the $k$th one is taken in the branch.
\item $N_i = \Sum b(\vy).P_{i+1}$
is negative, and the unique negative behaviour in $\bPsi_i$
is $\bN_{i_k} = \beta(\bQ_1,\ldots,$ $\bQ_s)$, so the rule $(\beta)$ is applied.
Among the upper sequents (recall that there is  one  sequent for each action
in $\beta_0$), the one
corresponding to $b(\vec{y}) \in \beta_0$ is taken in the branch.
\end{enumerate}
In this case, we define
$$
\cM(i)  :=  a(\vec{x}). z_{i_k}|\cob\langle \cM(y_1), \ldots, \cM(y_l)\rangle +\Sum_{\alpha_0 \setminus{\{a(\vec{x})\}}} c(\vec{w}).\maltese.$$
Here, the main additive component of $\cM(i)$ begins with
$a(\vec{x}). z_{i_k}|\cob$ because (1) $P_i$ begins with the positive action
$\coa$,
(2) the $k$th upper sequent is taken in the branch, and (3) the upper
sequent corresponding to $b(\vec{y})$ is taken.
The other additive components $\Sum_{\alpha_0 \setminus{\{a(\vec{x})\}}} c(\vec{w}).\maltese$ are needed
to ensure that our countermodel belongs to the
behaviour $\coal \langle \bN_1,\ldots, \bN_n \rangle\b$  (see Lemma \ref{l-comp} (1)).

The subdesigns $\cM(y_1), \ldots, \cM(y_l)$ are given by
$$\cM(y) :=
\Bigwedge \{ \cM(j) : \mbox{$P_j$ has head variable $y$}\}.$$
Notice that each $\cM(j)$ is a negative design, so the above
conjunction is a defined operation (in the sense of Definition \ref{conj extension} (2)).

We claim that $\cM(i)$ is well-defined,
because variables $\vy$ are chosen fresh, so do not
appear freely below $N_i \vdash \bPsi_i$. Hence
subdesigns $\cM(y_1), \ldots, \cM(y_l)$ do not have
$\cM(k)$ with $k \leq i$ as conjunct. Namely, $\cM(i)$ depends
only on $\cM(j)$ with $j> i$. This gives rise to
a recursive procedure and $\cM(i)$ arises in the limit of the procedure.

Notice also that the set $\{ \cM(j) : \mbox{$P_j$ has head variable $y$}\}$
can  be empty and  in such a case, we have
that $\cM(y) = \maltese^-$.

\begin{rem}
The above is an instance of corecursive definition. It is possible to
formally justify it by employing \emph{design generators} developed in
\cite{Terui08} (see in particular Theorem 2.12 of \cite{Terui08}).
An alternative way is
to define $\cM(i)$ (and $\cM(y)$) as
the limit of its finite approximations. Here we briefly outline
this latter approach.

We assume that $max = \infty$.
The idea is to chop off the branch $\mathsf{ob}$ at height $K$,
where $K$ is an arbitrary natural number, and
define finite approximations $\cM^K(i)$ and $\cM^K(y)$.
Then $\cM(i)$ and $\cM(y)$ arise as the limit when $K \rightarrow \infty$.

More concretely, given a natural number $K$,
we  define $\cM^K(i)$ by downward induction from $i=K$ to $i=0$
as follows:
\begin{enumerate}[$\bullet$]
\item When $i= K$, the sequent
$P_{K} \vdash \bTheta_{K}$ is of the form
$z|\coa\langle \vec{M}\rangle\vdash \bGamma, z: \coal \langle \vec{\bN}\rangle$.
We let $\cM^K(K) := \Sum_{a(\vx) \in \alpha_0} a(\vx).\maltese$.
\item When $i< K$, we proceed as in the case (iii) above. Namely,
\begin{eqnarray*}
\cM^K(i) & := &
a(\vec{x}). z_{i_k}|\cob\langle \cM^K(y_1), \ldots, \cM^K(y_l)\rangle +\Sum_{\alpha_0 \setminus{\{a(\vec{x})\}}} c(\vec{w}).\maltese, \\
\cM^K(y) & := &
\Bigwedge \{ \cM^K(j) : \mbox{$i< j \leq K$ and $P_j$ has head variable $y$}\},
\end{eqnarray*}
where actions $a(\vx)$, $\cob$ and the index $i_k$ are determined
as before.
\end{enumerate}

\noindent Now observe that the sequence $\{\cM^K(y)\}_{K\in \mathbb{N}}$
is ``monotone increasing'' in the sense
that $\cM^{K_2} (y)$ has more conjuncts than $\cM^{K_1}(y)$
whenever $K_1 < K_2$. The same for $\cM^{K}(i)$ with $i\leq K$.
Hence we can naturally obtain the ``limits''
$$
\cM(i) = \lim_{K\rightarrow\infty} \cM^K(i), \qquad
\cM(y) = \lim_{K\rightarrow\infty} \cM^K(y).
$$
This construction ends up with the same as the previous recursive one.
\end{rem}

Observe that each $\cM(i)$ and $\cM(x)$ thus constructed
are surely  models, \ie
atomic linear designs.
Theorem \ref{full completeness} is a direct consequence of
the following two lemmas.

The first lemma
crucially rests on
\emph{induction on logical behaviours}, that is an analogue
of  \emph{induction on formulas}, which lies at the core
of logical completeness in many cases.

\begin{lem}\label{l-comp}
For $P_i \vdash \bTheta_i$ appearing in the branch $\mathsf{ob}$ above, suppose that $P_i$ has a head variable $z$ and
$z:\bR \in \bTheta_i$. Then:
\begin{enumerate}[\em(1)]
\item  $\cM(i) \in \bR^\bot$;
\item $\cM(z)\in \bR^\bot$.
\end{enumerate}
\end{lem}

\proof
By induction on the construction of $\bR$.
\begin{enumerate}[(1)]
\item
Suppose that $i = max$. Since $\Omega$ does not have a head variable,
the case (i) does not apply. Hence we are in
the case (ii), namely $\bR =
\coal \langle \bN_1,\ldots,\bN_n\rangle$,
for some logical connective $\alpha$
and logical behaviours $\bN_1,\ldots,\bN_n$.
Thus,
$\bR\b = \alpha(\bN_1^\bot,\ldots,\bN_n^\bot)$, and
$\cM(max) := \Sum_{a(\vx) \in \alpha_0} a(\vx).\maltese$.

By internal completeness for negative connectives (Theorem \ref{t-negative}),
we have
$$ \begin{array}{ccl}
 \Sum a(\vx).P_a \in \alpha(\bN_1^\bot,\ldots, \bN_n^\bot)  & \Longleftrightarrow  &
 P_a  \models \vx : \vec{\bN}_a^\bot,
\mbox{ for every $a(\vx)\in\alpha_0$,}
\end{array}
$$
where $\vx= z_{i_1},\ldots,z_{i_m}$
and the expression $\vx : \vec{\bN}_a^\bot$ abbreviates the  positive context
$z_{i_1} : \bN_{i_1}^\bot,
\ldots, z_{i_m} : \bN_{i_m}^\bot$.
Since $\maltese \models \vx : \vec{\bN}_a^\bot$ trivially holds for
every $a(\vx)\in \alpha_0$, we have
$\cM(max) \in \alpha(\bN_1^\bot,\ldots, \bN_n^\bot) = \bR\b$.

When $i< max$, the case (iii) applies. In the same notation,
we have that
$\bR = \coal \langle \bN_1,\ldots,\bN_n \rangle$,
$\bN_{i_k} = \beta(\bQ_1,\ldots,$ $ \bQ_s)$, and
$$\cM(i) =  a(\vec{x}). z_{i_k}|\cob\langle \cM(y_{1}), \ldots,
\cM(y_{l})\rangle +\Sum_{\alpha_0 \setminus{\{a(\vec{x})\}}}c(\vec{w}).\maltese,$$
where actions $a(\vx)$, $\cob$, the index $i_k$ and the variables
$y_1, \dots, y_l$
are determined by the relevant
part of the branch $\mathsf{ob}$ as described above.

By induction hypothesis on (2), we have that
$\cM(y_{1}) \in \bQ_{1}^\bot, \ldots, \cM(y_{l}) \in \bQ_{l}^\bot$.
Hence,
$x_0|\cob \langle \cM(y_{1}), \ldots, \cM(y_{l})\rangle \in
\cobe \langle \bQ_1^\bot,\ldots,\bQ_s^\bot \rangle = \bN_{i_k}^\bot$.
Since
$\cM(y_{1}), \ldots, \cM(y_{l})$ are atomic (\ie closed),
we may derive
$z_{i_k}|\cob \langle \cM(y_{1}), \ldots, \cM(y_{l})\rangle \models
\vec{x} : \vec{\bN}_a^\bot$.
We also have $\maltese \models \vec{w}: \vec{\bN}_c^\bot$ for every
$c(\vec{w}) \in \alpha_0 \setminus{\{a(\vec{x})\}}$.
Hence, by internal completeness again,
$\cM(i) \in \alpha(\bN_1^\bot,\ldots, \bN_n^\bot) = \bR\b$.

\item It follows from (1) since $\bR\b$ is a negative logical behaviour
and so closed under $\Bigwedge$
(Theorem \ref{p-log} (4)).
\qed
\end{enumerate}

\medskip\noindent The proof of the next lemma suggests a similarity between
the construction of our countermodels and the \emph{B\"{o}hm-out}
technique (see, \eg \cite{Bar}), that constructs a suitable term context in order to visit a specific position in
the B\"ohm tree of a given $\lambda$-term.

Recall that the initial sequent of our open branch $\mathsf{ob}$ is
$P_0 \vdash \bTheta_0$ with
$\bTheta_0 = x_1: \bP_1, \dots, x_n:\bP_n$, so that
$\mathsf{fv}(P_0) \subseteq \{x_1, \dots, x_n\}$. We have:
\begin{lem}\label{l-bohm}
$$\Norm{P_0[\cM(x_1)/x_1, \dots, \cM(x_n)/x_n]} = \Omega.$$
\end{lem}

\proof
We first prove that there is a reduction sequence
$$P_i [\cM(v_1)/v_1, \dots, \cM(v_s)/v_s] \nondet^*
P_{i+1} [\cM(w_1)/w_1, \dots, \cM(w_t)/w_t]$$
for
any $i< max$, where $v_1,\ldots,v_s$ and $w_1,\ldots,w_t$ are the free variables of
$P_i$ and $P_{i+1}$, respectively.
Suppose that $P_i$ is as in the case (iii) above, so has the head variable
$z \in \{v_1,\ldots,v_s\}$.
By writing $[\theta]$ for $[\cM(v_1)/v_1, \dots, \cM(v_s)/v_s]$
and noting that
$\cM(z)$  is a (defined) conjunction that contains $\cM(i) =
a(\vec{x}). z_{i_k}|\cob \langle \cM(y_{1}), \ldots, \cM(y_{l})\rangle
+\Sum_{\alpha_0 \setminus{\{a(\vec{x})\}}}c(\vec{w}).\maltese$ as conjunct,
we have:
$$
\begin{array}{rcl}
P_i [\theta] & = &
\cM(z) \ | \ \coa\langle M_1[\theta],\ldots,M_{{k-1}}[\theta], \Sum b(\vy).P_{i+1}[\theta], M_{k+1}[\theta],\ldots,M_{{m}}[\theta]\rangle  \\
& \nondet & \left(\Sum b(\vy).P_{i+1}[\theta]\right) \;
|\;  \cob \langle \cM(y_{1}), \ldots, \cM(y_{l})\rangle  \; \wedge \cdots\\
&  \nondet &  P_{i+1}[\theta, \cM(y_{1})/ y_{1}, \ldots, \cM(y_{l})/y_{l}],\\
\end{array}
$$
as desired.
When $max = \infty$, we have obtained an infinite reduction sequence
from $P_0[\cM(x_1)/x_1, \dots, \cM(x_n)/x_n]$.
Otherwise,
$P_0[\cM(x_1)/x_1, \dots, \cM(x_n)/x_n] \nondet^* P_{max}[\theta]$,
for some substitution $[\theta]$.

In case (i), we have
$P_{max} = P_{max}[\theta] = \Omega$, while
in case (ii), we have $P_{max} = z|\coc\langle \vec{M}\rangle$. So,
$$P_{max}[\theta] =
z|\coc\langle \vec{M}\rangle [\theta] =
\cM(z)|\coc\langle \vec{M}[\theta]\rangle
\nondet\Omega,$$
because $\cM(z)$ contains
$\cM(max)$ as conjunct, and $\cM(max)
 = \Sum_{a(\vx) \in \alpha_0} a(\vec{x}).\maltese$
has
$c(\vec{w}).\Omega$ as component.
\qed

Theorem \ref{full completeness} now follows easily. Suppose that $P_0 \vdash x_1:\bP_1, \dots, x_n: \bP_n$
is not derivable. Then we obtain models $\cM(x_1) \in \bP_1^\bot$, \dots, $\cM(x_n) \in \bP_n^\bot$
by Lemma \ref{l-comp} and $\Norm{P_0[\cM(x_1)/x_1, \dots, \cM(x_n)/x_n]}=\Omega$ by Lemma \ref{l-bohm}.
This means that $P_0 \not\models x_1:\bP_1, \dots, x_n: \bP_n$.

Our explicit construction
of the countermodels yields a by-product:
\begin{cor}[Downward L\"owenheim-Skolem, Finite model property] \label{lowen} \hfill
\begin{enumerate}[\em(1)]
\item Let $P$ be a proof and $\bP$ a logical behaviour.
If $P \not\in \bP$, then there is a countable model $M \in \bP^\bot$
(\ie $\mathsf{ac}^+(M)$ is a countable set)
such that $P\nv M$.
\item
Furthermore, when $P$ is linear,
there is a finite and deterministic model $M \in \bP^\bot$ such that $P\nv M$. \qed
\end{enumerate}
\end{cor}

\noindent The second statement is due to the observation that when $P$ is linear
the positive rule $(\coal, \coa)$ can be replaced with a linear variant:
$$
\infer[(\coal,\coa)_{lin}]{z|\coa\langle M_1, \dots, M_m
\rangle\vdash
\bGamma, z:\coal\langle\bN_1, \dots, \bN_n\rangle}{
M_1 \vdash \bGamma_1, \bN_{i_1} &
\dots &
M_m \vdash \bGamma_m, \bN_{i_m}},
$$
where $\bGamma_1, \dots, \bGamma_m$ are disjoint subsets of $\bGamma$.
We then immediately see that the proof search tree is always finite,
and so is the model $\cM(x)$. It is deterministic, since
each variable occurs at most once as head variable in a branch
so that all conjunctions are at most unary.

\section{Conclusion and related work}\label{s-conclusion}
\noindent We have presented
a G\"odel-like completeness theorem for proofs in the framework
of ludics,
aiming at linking completeness theorems for provability
with those for proofs. We have explicitly constructed a
countermodel against any failed proof attempt, following
Sch\"utte's idea based on cut-free proof search.
Our proof employs K\"onig's lemma and reveals a sharp opposition
between finite proofs and infinite models, leading to a clear
analogy with L\"owenhein-Skolem theorem.
Our proof also employs an analogue of the B\"ohm-out technique
\cite{Bohm,Bar}
(see the proof of Lemma \ref{l-bohm}), though it does not lead to the
separation property (Remark \ref{r-separation}).

In Hyland-Ong game semantics, Player's innocent strategies
most naturally correspond to
possibly infinite B\"ohm trees (see, \eg \cite{Curiennote}).
One could of course impose
finiteness (or compactness) on them to have correspondence
with finite proofs.
 But it would not lead to an explicit
construction of Opponent's strategies defeating infinite
proof attempts. Although finiteness is
imposed in \cite{Basaldella09} too, our current work shows
that it is not necessary in ludics.

Our work also highlights the duality:
$$
\begin{array}{ccc}
\mbox{\textbf{proof}} & \rightleftharpoons & \mbox{\textbf{model}} \\
\mbox{\emph{deterministic, nonlinear}} &  & \mbox{\emph{nondeterministic, linear}}\\
\end{array}
$$
The principle is that \emph{when proofs admit contraction, models
have to be nondeterministic} (whereas they do not have to be nonlinear).

A similar situation
arises in some variants of $\lambda$-calculus
and linear logic,
when one proves the
separation property.

We mention
 \cite{Dezani-Intrigila-Venturini:ICTCS-98}, where the authors add
 a  nondeterministic choice operator and a numeral
system    to the pure $\lambda$-calculus
in order  to  internally  (interactively)  discriminate
two pure $\lambda$-terms that have different B\"ohm trees.
However, in contrast to our work, the nondeterminism   needed for their purpose is of existential nature:
a term converges if at least one of the possible reduction
sequences starting from it terminates.

In
\cite{DBLP:conf/lpar/MazzaP07},
the separation property for
differential interaction nets \cite{DBLP:journals/tcs/EhrhardR06}
 is proven.
A key point is that
 the exponential modalities  in differential interaction nets are more
``symmetrical'' than in linear logic.
In our setting, the symmetry shows up between
nonlinearity and nondeterministic conjunctions (\ie nonuniform elements).
It is  typically found in Theorem \ref{p-duphalf}, which reveals a tight
connection between duplicability of positive logical behaviours
and closure under nondeterministic conjunctions of negative logical behaviours.
Similar   nonuniform  structures
naturally arise in various  semantical models based on coherence spaces and games, such as
finiteness spaces \cite{Ehr},
 indexed linear logic and nonuniform coherence spaces \cite{BucEhr},
 nonuniform hypercoherences \cite{Boudes},
and asynchronous games \cite{Mellies} (see also \cite{Basaldella09}).

%
%
%
%


For future work,
we plan to extend
 our setting by enriching
the proof system with propositional variables, second order quantifiers and nonlogical axioms.
By moving to the second order setting,
we hope to give an \emph{interactive} account to G\"odel's incompleteness
theorems as well.

\section*{Acknowledgement}
\noindent We are deeply indebted to Pierre-Louis Curien, who
gave us a lot of useful comments.
Our thanks are also
due to the anonymous referees.


\appendix

\section{Correspondence with polarized linear logic} \label{pol LL}

\noindent In this appendix, we show a correspondence between
the proof system for ludics introduced in \ref{proof system}
and the constant-only propositional fragment of  \emph{polarized linear logic}
$\mathbf{LLP}$  \cite{DBLP:journals/apal/Laurent04}. This will ensure that our proof system
is rich enough to capture a constructive variant of constant-only propositional classical logic.


\subsection{Syntax of $\mathbf{LLP}$}
We recall the syntax of the constant-only
propositional  fragment of $\mathbf{LLP}$. The \emph{formulas} are split into positive and negative
ones and generated by the following grammar:
$$
\begin{array}{ccccccccccc}
\sP  & ::= & \mathsf{0} & | & \mathsf{1} & | & \sP \otimes \sP & | & \sP \oplus \sP & | & \oc \sN, \\
\sN  & ::= & \mathsf{\top} & | & \mathsf{\bot} & | & \sN \parr \sN & | & \sN \with \sN & | & \wn \sP. \\
\end{array}
$$
The linear negation is defined in the usual way.
A \emph{sequent} is of the form $\vdash \sGamma$ with $\sGamma$ a multiset of formulas.
The inference rules of $\mathbf{LLP}$ are given below:
\begin{center}
\begin{tabular}{cccc}

\AxiomC{}
\UnaryInfC{$\vdash \mathsf{\Gamma},\mathsf{\top}$} \DisplayProof

&
\AxiomC{}
\UnaryInfC{$\vdash\mathsf{1}$} \DisplayProof

&
\AxiomC{$\vdash \mathsf{\Gamma}$}
\UnaryInfC{$\vdash \mathsf{\Gamma},\mathsf{\bot}$} \DisplayProof

&

\AxiomC{$\vdash \mathsf{\Gamma}, \sP $}
\AxiomC{$\vdash \mathsf{\Delta}, \sQ $}
\BinaryInfC{$\vdash \mathsf{\Gamma},\mathsf{\Delta}, \sP \otimes \sQ $} \DisplayProof

\\

&&&\\

\AxiomC{$\vdash \mathsf{\Gamma}, \sP_i$}
\UnaryInfC{$\vdash \mathsf{\Gamma}, \sP_1 \oplus \sP_2$} \DisplayProof

&

\AxiomC{$\vdash \mathsf{\Gamma}, \sN,\sM $}
\UnaryInfC{$\vdash \mathsf{\Gamma}, \sN \parr \sM $} \DisplayProof

&

\AxiomC{$\vdash \mathsf{\Gamma}, \sN $}
\AxiomC{$\vdash \mathsf{\Gamma}, \sM $}
\BinaryInfC{$\vdash \mathsf{\Gamma}, \sN \with \sM $} \DisplayProof

&

\AxiomC{$\vdash \mathsf{\cN},  \sN$}
\UnaryInfC{$\vdash \mathsf{\cN}, \oc \sN$} \DisplayProof\\

&&&\\

\AxiomC{$\vdash \mathsf{\Gamma}, \sP $}
\UnaryInfC{$\vdash \mathsf{\Gamma}, \wn \sP $} \DisplayProof

&

\AxiomC{$\vdash \mathsf{\Gamma}  $}
\UnaryInfC{$\vdash \mathsf{\Gamma},\sN  $} \DisplayProof
&
\AxiomC{$\vdash \mathsf{\Gamma},\sN,\sN  $}
\UnaryInfC{$\vdash \mathsf{\Gamma},\sN  $} \DisplayProof

&

\AxiomC{$\vdash \mathsf{\Gamma}, \sN $}
\AxiomC{$\vdash \mathsf{\Delta}, \sN\b $}
\BinaryInfC{$\vdash \mathsf{\Gamma},\mathsf{\Delta} $} \DisplayProof

\\

\end{tabular}
 \end{center}
 where:
 \begin{enumerate}[$\bullet$]
\item in the $\mathsf{\top}$-rule above $\mathsf{\Gamma}$ contains at most one
positive formula;
\item   $\mathsf{\cN}$ denotes
a context consisting of negative formulas only.
\end{enumerate}
In  \cite{LaurentThesis} it is proven that
if $\vdash \mathsf{\Gamma}$ is provable in $\mathbf{LLP}$,
then $\mathsf{\Gamma}$  contains at most one positive formula.
Notice that it is strictly opposite to the ludics discipline
\cite{DBLP:journals/mscs/Girard01};
in the latter, any sequent contains at most one \emph{negative} behaviour.
To resolve this mismatch, we modify $\mathbf{LLP}$ in several steps,
making it closer to the ludics discipline.

Precisely, in Section \ref{strict derivations}
we introduce the concept of strict sequent
which leads us to the formulation
of syntectic connectives in
$\mathbf{LLP}$   (Section \ref{synth llp}).
In Section \ref{ludics to LLP}, we give an embedding of
$\mathbf{LLP}$  with synthetic connectives
into the proof
system of ludics we gave in Section
\ref{proof system}.
Finally, in Section \ref{llp to ludics} we give a converse
embedding of the ludics proof system into $\mathbf{LLP}$.

\subsection{Restriction to strict derivations.} \label{strict derivations}
We call a sequent of $\mathbf{LLP}$ \emph{strict} if it is of the form
$\vdash \wn\sGamma, \sD$, where $\sD$ is an arbitrary formula.
In particular, $\vdash \wn\sGamma$ is strict.
We modify the inference rules as follows:
\begin{enumerate}[$\bullet$]
\item Structural rules are made implicit by absorbing weakening and contraction
into logical inference rules.
\item The rules for positive connectives and the $\wn$-dereliction rule are
restricted to strict sequents.
\item The cut rule is omitted.
\end{enumerate}

\medskip\noindent We thus obtain the following inference rules:
\begin{center}
\begin{tabular}{cccc}
\AxiomC{}
\UnaryInfC{$\vdash \mathsf{\Gamma},\mathsf{\top}$} \DisplayProof
 &
\AxiomC{}
\UnaryInfC{$\vdash\wn\sGamma,\mathsf{1}$} \DisplayProof
 &
\AxiomC{$\vdash \mathsf{\Gamma}$}
\UnaryInfC{$\vdash \mathsf{\Gamma},\mathsf{\bot}$} \DisplayProof
 &
\AxiomC{$\vdash \wn\mathsf{\Gamma}, \sP $}
\AxiomC{$\vdash \wn\mathsf{\Gamma}, \sQ $}
\BinaryInfC{$\vdash \wn\mathsf{\Gamma}, \sP \otimes \sQ $} \DisplayProof
\\[2em]
\AxiomC{$\vdash \wn\mathsf{\Gamma}, \sP_i$}
\UnaryInfC{$\vdash \wn\mathsf{\Gamma}, \sP_1 \oplus \sP_2$} \DisplayProof
&
\AxiomC{$\vdash \mathsf{\Gamma}, \sN,\sM $}
\UnaryInfC{$\vdash \mathsf{\Gamma}, \sN \parr \sM $} \DisplayProof
&
\AxiomC{$\vdash \mathsf{\Gamma}, \sN $}
\AxiomC{$\vdash \mathsf{\Gamma}, \sM $}
\BinaryInfC{$\vdash \mathsf{\Gamma}, \sN \with \sM $} \DisplayProof
&
\AxiomC{$\vdash \wn\sGamma,  \sN$}
\UnaryInfC{$\vdash \wn\sGamma, \oc \sN$} \DisplayProof
\\[2em]
\AxiomC{$\vdash \wn\sGamma, \sP $}
\AxiomC{$(\wn\sP \in \, \wn\sGamma)$}
\BinaryInfC{$\vdash \wn\sGamma$} \DisplayProof
\end{tabular}\\
\end{center}
We call the resulting proof system $\mathbf{LLP}_{str}$.

Notice that a derivation of a strict sequent in
$\mathbf{LLP}_{str}$ may involve sequents which are
not strict.  For instance, consider:
\begin{center}
\AxiomC{}
\UnaryInfC{$\vdash \wn\sGamma, \mathsf{\top}$}
\UnaryInfC{$\vdash \wn\sGamma, \mathsf{\top},\mathsf{\bot}$}
\UnaryInfC{$\vdash \wn\sGamma, \mathsf{\top} \parr \mathsf{\bot}$} \DisplayProof
\end{center}

\medskip\noindent The following property can be easily verified by
taking into account the invertibility of negative rules and the
focalization property of positive rules \cite{Andreoli}.

\begin{lem}
A strict sequent is provable in $\mathbf{LLP}$
if and only if it is provable in $\mathbf{LLP}_{str}$. \qed
\end{lem}

\noindent Strict sequents will play a crucial role for the correspondence
between $\mathbf{LLP}$ and the proof system for ludics
(Theorem \ref{t-llp}).
The intuition, which we will formalize later, is that a strict sequent
$\vdash \wn\sP_1,\ldots,\wn\sP_n, \sD $ can be
thought of as a  sequent of the proof system of ludics
(omitting the information about designs)
of the form
$\vdash \bP_1,\ldots, \bP_n, \bD$.

On the other hand, strict derivations
serve as intermediate step to define  synthetic connectives
and the proof system $\mathbf{LLP}_{syn}$ we give in the next
section.

\subsection{Synthetic connectives.} \label{synth llp}
Any derivation of a strict sequent in $\mathbf{LLP}_{str}$ can be decomposed into subderivations of the following forms:
\begin{enumerate}[(i)]
\item Positive subderivation:
$$
\infer*{\vdash \wn\sGamma, \sP(\sN_1, \dots, \sN_n)}{
\vdash \wn\sGamma, \sN_{i_1} &
\ \,\,\! \cdots &
\vdash \wn\sGamma, \sN_{i_m}}
$$
that consists of positive inference rules only,
where $\sP(\sN_1, \dots, \sN_n)$ is a positive formula obtained from $\sN_1, \dots, \sN_n$ by
applying positive connectives, and $i_1, \dots, i_m \in \{1, \dots, n\}$.
For instance, if
$\sP(\sN_1, \dots, \sN_n)$ is of the form
$\mathsf{1}\otimes (\oc\sN \otimes (\oc\sM
\oplus \oc\sL))$, there are two positive subderivations
with conclusion $\vdash \wn\sGamma,
\mathsf{1}\otimes (\oc\sN \otimes (\oc\sM
\oplus \oc\sL))$:

$$
\infer{\vdash \wn\sGamma, \mathsf{1}\otimes (\oc\sN \otimes (\oc\sM \oplus \oc\sL))}{
\infer{\vdash \wn\sGamma, \mathsf{1}}{}
&
\infer{\vdash \wn\sGamma, \oc\sN \otimes (\oc\sM \oplus \oc\sL)}{
\infer{\vdash \wn\sGamma, \oc\sN}{\vdash \wn\sGamma, \sN}
&
\infer{\vdash \wn\sGamma, \oc\sM \oplus \oc\sL}{
\infer{\vdash \wn\sGamma, \oc\sM}{\vdash \wn\sGamma, \sM}
}}}
\qquad
\infer{\vdash \wn\sGamma, \mathsf{1}\otimes (\oc\sN \otimes (\oc\sM \oplus \oc\sL))}{
\infer{\vdash \wn\sGamma, \mathsf{1}}{}
&
\infer{\vdash \wn\sGamma, \oc\sN \otimes (\oc\sM \oplus \oc\sL)}{
\infer{\vdash \wn\sGamma, \oc\sN}{\vdash \wn\sGamma, \sN}
&
\infer{\vdash \wn\sGamma, \oc\sM \oplus \oc\sL}{
\infer{\vdash \wn\sGamma, \oc\sL}{\vdash \wn\sGamma, \sL}
}}}
$$
\item Negative subderivation:
$$
\infer*{\vdash \wn\sGamma, \sN(\sP_1, \dots, \sP_n)}{
\vdash \wn\sGamma,  \wn\vec{\sP}_1 & \, \cdots &
\vdash \wn\sGamma,  \wn\vec{\sP}_k}
$$
that consists of negative inference rules only,
where $\sN(\sP_1, \dots, \sP_n)$
is a negative formula obtained from $\sP_1, \dots, \sP_n$ by
applying negative connectives, and $ \vec{\sP}_1, \dots, \vec{\sP}_k$ consist of
formulas in $\{\sP_1, \dots, \sP_n\}$.
For instance, if $\sN(\sP_1, \dots, \sP_n)$
is of the form
$\mathsf{\bot}\parr (\wn\sP \parr (\wn\sQ \with \wn\sR))$,
then there is (essentially) one negative subderivation
with conclusion
 $\vdash \wn\sGamma, \mathsf{\bot}\parr (\wn\sP \parr (\wn\sQ \with \wn\sR))$:

$$
\infer{\vdash \wn\sGamma, \mathsf{\bot}\parr (\wn\sP \parr (\wn\sQ \with \wn\sR))}{
\infer{\vdash \wn\sGamma, \mathsf{\bot}, \wn\sP \parr (\wn\sQ \with \wn\sR)}{
\infer{\vdash \wn\sGamma,  \wn\sP \parr (\wn\sQ \with \wn\sR)}{
\infer{\vdash \wn\sGamma,  \wn\sP, \wn\sQ \with \wn\sR}{
\vdash \wn\sGamma, \wn\sP, \wn\sQ & \vdash \wn\sGamma, \wn\sP, \wn\sR}}}}
$$
\item $\wn$-dereliction:
$$
\infer{\vdash \wn\sGamma}{\vdash \wn\sGamma, \sP & (\wn\sP \in \, \wn\sGamma)}
$$
\end{enumerate}

\medskip\noindent Notice that the premises and conclusion of each subderivation
are assumed to be strict sequents; one can easily check that it is always the case in
any derivation of a strict sequent in $\mathbf{LLP}_{str}$.

The above decomposition motivates us to cluster the logical connectives of the same polarity into
 \emph{synthetic connectives}
(\cf \cite{Girard00}). Consider the expressions finitely generated by:
$$
\begin{array}{ccccccccccc}
\sPl  & ::= & \mathsf{0} & | & \mathsf{1} & | & \sPl \otimes \sPl & | & \sPl \oplus \sPl & | & \oc x, \\
\sNl  & ::= & \mathsf{\top} & | & \mathsf{\bot} & | & \sNl \parr \sNl & | & \sNl \with \sNl & | & \wn x,
\end{array}
$$
where $x$ ranges over the set of variables.

We write $\mathsf{var}(\sPl)$ (resp.\
$\mathsf{var}(\sNl)$)
to denote the set of variables occurring in
$\sPl$ (resp.\ $\sNl$).
$\sPl$ is a
\emph{positive synthetic connective}
if for every subexpression of $\sPl$ of the form $\sPl_1 \otimes \sPl_2$,
$\mathsf{var}(\sPl_1)$ and $\mathsf{var}(\sPl_2)$ are disjoint.
For instance, $\oc x \otimes (\oc y \oplus \oc y)$ is a positive synthetic connective
while $\oc x \oplus (\oc y \otimes \oc y)$ is not.
Likewise, $\sNl$ is a
\emph{negative synthetic connective}
if for every subexpression of $\sNl$ of the form $\sNl_1 \parr \sNl_2$,
$\mathsf{var}(\sNl_1)$ and $\mathsf{var}(\sNl_2)$ are disjoint.
This condition is needed when we translate
synthetic connectives to logical connectives of ludics.

We indicate the variables occurring in $\sPl$ by writing $\sPl = \sPl(x_1, \dots, x_n)$,
and similarly for $\sNl$.
Given a negative synthetic connective $\sNl$, its \emph{dual}
$\sNl^d$ is obtained by replacing $\mathsf{\top}$ with $\mathsf{0}$,
$\bot$ with $\mathsf{1}$,
$\parr$ with $\otimes$, $\with$ with $\oplus$, and $\wn$ with $\oc$
respectively, in each occurrence of symbol. $\sPl^d$ is similarly defined.\\

The \emph{formulas} of $\mathbf{LLP}$ are then redefined inductively as follows:
\begin{eqnarray*}
\sP & ::= & \sPl(\sN_1, \dots, \sN_n), \\
\sN & ::= & \sNl(\sP_1, \dots, \sP_n),
\end{eqnarray*}
where $\sPl(\sN_1, \dots, \sN_n)$ is obtained from $\sPl = \sPl(x_1, \dots, x_n)$
by substituting $\sN_i$ for $x_i$ ($1\leq i\leq n$).
Notice that when $n=0$,  $\sP$ can be any
combination of
$\mathsf{0}$ and  $\mathsf{1}$ using $\otimes$ and $\oplus$.

To each positive synthetic connective
$\sPl(x_1, \dots, x_n)$, we can naturally associate
a set of inference rules as follows. Consider all possible
positive subderivations with conclusion
$\vdash \wn\sGamma, \sPl(\sN_1, \dots, \sN_n)$ in the sense of (i) above.
To each such derivation
$$
\infer*{\vdash \wn\sGamma, \sP(\sN_1, \dots, \sN_n)}{
\vdash \wn\sGamma, \sN_{i_1} &
\ \,\,\! \cdots &
\vdash \wn\sGamma, \sN_{i_m}}
$$
we associate an inference rule:
$$
\infer{\vdash \wn\sGamma, \sP(\sN_1, \dots, \sN_n)}{
\vdash \wn\sGamma, \sN_{i_1} &
\ldots &
\vdash \wn\sGamma, \sN_{i_m}}
$$
For instance,
to $\sPl(x, y,z) = \mathsf{1}\otimes (\oc x \otimes (\oc y \oplus \oc z))$,
we associate two inference rules:

$$
\infer{\vdash \wn\sGamma, \sPl(\sN, \sM, \sL)}{
\vdash \wn\sGamma, \sN &
\vdash \wn\sGamma, \sM}
\qquad
\infer{\vdash \wn\sGamma, \sPl(\sN, \sM, \sL)}{
\vdash \wn\sGamma, \sN &
\vdash \wn\sGamma, \sL}
$$

Likewise, each negative synthetic connective
$\sNl(x_1, \dots, x_n)$
comes equipped with a unique inference rule
derived from the negative subderivation with conclusion
$\vdash \wn\sGamma, \sNl(\sP_1, \dots, \sP_n)$
(see (ii) above).
For instance, $\sNl(x,y,z) = \mathsf{\bot}\parr (\wn x \parr (\wn y \with \wn z))$ is equipped with:

$$
\infer{\vdash \wn\sGamma, \sNl(\sP, \sQ, \sR)}{
\vdash \wn\sGamma, \wn\sP, \wn\sQ & \vdash \wn\sGamma, \wn\sP, \wn\sR}
$$

Observe the asymmetry between the positive and negative cases here;
in the negative case,
we leave the $\wn$-formulas $\wn\sP, \wn\sQ, \wn\sR$ in the premises.
These formulas are to be dealt with
by the $\wn$-dereliction rule.

We thus consider proof system $\mathbf{LLP}_{syn}$ that consists of three types of inference rules:
$$
\infer{\vdash \wn\sGamma, \sPl(\sN_1, \dots, \sN_n)}{
\vdash \wn\sGamma, \sN_{i_1} & \dots & \vdash \wn\sGamma, \sN_{i_m}}
\qquad
\infer{\vdash \wn\sGamma, \sNl(\sP_1, \dots, \sP_n)}{
\vdash \wn\sGamma,  \wn\vec{\sP}_1 & \dots &
\vdash \wn\sGamma,  \wn\vec{\sP}_k}
\qquad
\infer{\vdash \wn\sGamma}{\vdash \wn\sGamma, \sP & (\wn\sP \in \wn\sGamma)}
$$
In view of the decomposition of $\mathbf{LLP}_{str}$ derivations,
we obviously have:
\begin{lem}
A strict sequent is provable in $\mathbf{LLP}_{str}$
if and only if it is provable in $\mathbf{LLP}_{syn}$. \qed
\end{lem}

\subsection{Relating to the ludics proof system.}
\label{ludics to LLP}

Let us now move on to the proof system for ludics described in \ref{proof system}.
We assume that the signature $\mathcal{A}$ is rich enough to interpret $\mathbf{LLP}$:
\begin{enumerate}[$\bullet$]
\item $\mathcal{A}$ contains a nullary name $*$ and a unary name $\uparrow$.
\item If $\mathcal{A}$ contains an $n$-ary name $a$ and an $m$-ary name $b$,
it also contains $n$-ary names $\pi_1 a$, $\pi_2 a$ and an $(n+m)$-ary name $ a \wp b$
(\cf Example \ref{linear}).
\end{enumerate}

\medskip \noindent Given a negative synthetic connective $\sNl$, we
inductively associate a set $\sNl^\bullet_0$ of negative actions of
ludics as follows:
\begin{eqnarray*}
\mathsf{\top}^\bullet_0 & = & \emptyset, \\
\mathsf{\bot}^\bullet_0 & = & \{*\},\\
\mathsf{\wn}x^\bullet_0 & = & \{\uparrowd(x)\},\\
(\sNl \parr \sMl)^\bullet_0 & = &  \{ a \wp b (\vx, \vy) : a(\vx) \in \sNl^\bullet_0, b(\vy) \in \sMl^\bullet_0\}, \\
(\sNl \with \sMl)^\bullet_0 & = &  \{ \pi_1 a (\vx) : a(\vx) \in \sNl^\bullet_0\}\cup \{ \pi_2 b (\vy) : b(\vy) \in \sMl^\bullet_0\}.
\end{eqnarray*}
Notice that when
$a(\vx) \in \sNl^\bullet_0$,  the variables $\vx$ occur in $\sNl$. Hence
$a \wp b (\vx, \vy)$ above is certainly a negative action,
since $\vx$ and $\vy$ are disjoint sequences
due to the definition of
negative synthetic connective.
We finally let $\sNl^\bullet = (\vz, \sNl^\bullet_0)$, where
$\vz$ lists the variables occurring in $\sNl$.
A positive synthetic connective $\sPl$ is interpreted by
$\sPl^\bullet = \sPl^{d\bullet}$.

For instance, when
$\sPl(x, y,z) = \mathsf{1}\otimes (\oc x \otimes (\oc y \oplus \oc z))$
and
$\sNl(x,y,z) = \mathsf{\bot}\parr (\wn x \parr (\wn y \with \wn z))$,
we have $\sPl^\bullet = \sNl^\bullet = (x,y,z, \sNl^\bullet_0)$ with
$$
\sNl^\bullet_0 = \{*\wp (\uparrowd\wp(\pi_1\uparrowd))(x,y),\
*\wp (\uparrowd\wp(\pi_2\uparrowd))(x,z)\}.$$
This induces a polarity-preserving translation from the formulas of $\mathbf{LLP}$ to the logical behaviours of ludics:
\begin{eqnarray*}
\sNl(\sP_1, \dots, \sP_n)^\bullet & := & \sNl^\bullet (\sP_1^\bullet, \dots, \sP_n^\bullet), \\
\sPl(\sN_1, \dots, \sN_n)^\bullet & := & \overline{\sPl^{\bullet}} \langle \sN_1^\bullet, \dots, \sN_n^\bullet \rangle.
\end{eqnarray*}

\medskip\noindent To establish a connection with $\mathbf{LLP}$,
we simplify the proof system of \ref{proof system} by taking its \emph{skeleton}, namely by
omitting all information about designs.
The resulting proof system, which we call $\mathbf{L}$, consists of two sorts of inference rules:
$$
\infer[(\coal, \coa)]{\vdash \bGamma}{
 \vdash \bGamma, \bN_{i_1} &
\dots &
 \vdash \bGamma, \bN_{i_m}
& (\coal\langle\bN_1, \dots, \bN_n\rangle \in \bGamma)}
\qquad
\infer[(\alpha)]{ \vdash
\bGamma, \alpha(\bP_1, \dots, \bP_n)}{
\{\vdash \bGamma,  \bP_{i_1}, \dots,  \bP_{i_m}\}_{a(\vx)\in \alpha_0}}
$$
where  $\alpha = (\vz, \alpha_0)$, $\vz = z_1, \dots, z_n$, $a(\vx) \in \alpha_0$ and
the indices
$i_1, \dots, i_m \in \{1, \dots,  n\}$ are determined by the variables
$\vec{x} = z_{i_1}, \dots, z_{i_m}$.

For instance, when
$\sPl(x, y,z) = \mathsf{1}\otimes (\oc x \otimes (\oc y \oplus \oc z))$ and
$\sNl(x,y,z) = \mathsf{\bot}\parr (\wn x \parr (\wn y \with \wn z))$,
we have the following inference rules for $\sPl^\bullet$ and
$\sNl^\bullet$:
$$
\infer{\vdash\bGamma}{\vdash \bGamma, \bN & \vdash \bGamma, \bM
& (\overline{\sPl^\bullet} \langle \bN, \bM, \bL\rangle \in \bGamma)}
\qquad
\infer{\vdash\bGamma}{\vdash \bGamma, \bN & \vdash \bGamma, \bL
& (\overline{\sPl^\bullet} \langle \bN, \bM, \bL\rangle \in \bGamma)}
$$
$$
\infer{\vdash\bGamma,\sNl^\bullet (\bP, \bQ, \bR)}{
\vdash\bGamma, \bP, \bQ &
\vdash\bGamma, \bP, \bR}
$$
It is now straightforward to verify:
\begin{lem}
A strict sequent $\vdash \wn\sGamma, \sD$ is derivable in $\mathbf{LLP}_{syn}$ if and only if
$\vdash \sGamma^\bullet, \sD^\bullet$ is derivable in $\mathbf{L}$. \qed
\end{lem}

\noindent We therefore obtain:
\begin{thm}\label{t-llp}
A strict sequent $\vdash \wn\sGamma, \sD$ is derivable in $\mathbf{LLP}$ if and only if
$\vdash \sGamma^\bullet, \sD^\bullet$ is derivable in $\mathbf{L}$. \qed
\end{thm}

One can annotate derivations in $\mathbf{L}$ with designs as in Section \ref{proof system}. Therefore the above theorem means that
ludics designs can be used as term syntax for $\mathbf{LLP}$, as far as strict
sequents and derivations are concerned (although we have to verify carefully that the translation
preserves the reduction relation).

\subsection{From ludics to $\mathbf{LLP}$} \label{llp to ludics}
It is also possible to give a converse translation from the logical behaviours of ludics to the formulas of $\mathbf{LLP}$.
To do so, we proceed as follows (\cf Example \ref{example}):
\begin{enumerate}[$\bullet$]
\item to each action $a(x_1, \dots, x_m)$, we associate the synthetic connective
$a(x_1, \dots, x_m)^\circ := \wn x_1 \parr \dots \parr \wn x_m$
($a()^\circ := \mathbf{\bot}$, if $a$ is nullary);
\item to each logical connective $\alpha = (\vz, \{a_1(\vx_1), \dots, a_k(\vx_k)\})$, we associate the synthetic connective
$\alpha^\circ := a_1(\vx_1)^\circ \with \dots \with a_k(\vx_k)^\circ$
($\alpha^\circ := \mathbf{\top}$, if $k=0$);
\item to each logical behaviour, we associate the formula
of $\mathbf{LLP}$
\begin{eqnarray*}
\alpha(\bP_1, \dots, \bP_n)^\circ & := & \alpha^\circ (\bP_1^\circ, \dots, \bP_n^\circ), \\
\coal\langle \bN_1,\ldots,\bN_n \rangle^\circ & := & \alpha^{\circ d}(\bN_1^\circ, \dots, \bN_n^\circ);
\end{eqnarray*}
\item to each positive context $\bGamma = \bP_1, \dots, \bP_n$ of system $\mathbf{L}$, we associate the multiset
 $\bGamma^\circ := \wn\bP^\circ_1, \dots, \wn\bP^\circ_n$ of formulas of $\mathbf{LLP}$;
\item  to each negative context $\bGamma, \bN$ of system $\mathbf{L}$,
we associate the multiset of formulas  $(\bGamma, \bN)^\circ := \bGamma^\circ, \bN^\circ$.
\end{enumerate}

\medskip\noindent It is routine to define an isomorphism between $\sD$ and $\sD^{\bullet\circ}$ (resp.\ between $\bD$ and $\bD^{\circ\bullet}$)
in some natural sense. Moreover, the translation of a sequent of system $\mathbf{L}$
always results in a strict sequent of $\mathbf{LLP}$.
We therefore conclude by Theorem \ref{t-llp}:
\begin{thm}
A sequent $\vdash \bLambda$ is derivable in proof system $\bL$ if and only if
$\vdash \bLambda^\circ$ is derivable in $\mathbf{LLP}$. \qed
\end{thm}

\end{document}